\documentclass[journal]{IEEEtran}
\ifCLASSINFOpdf
\else

\fi

\usepackage{ifpdf}
\usepackage[pdftex]{graphicx}
\usepackage[cmex10]{amsmath}
\usepackage{subfig}
\usepackage{amssymb}
\usepackage{mathrsfs}
\usepackage{amsmath}
\usepackage{indentfirst}
\usepackage{booktabs}
\usepackage{multirow}
\usepackage{graphicx}
\usepackage{epstopdf}
\usepackage{fixltx2e}
\usepackage{soul}
\usepackage{units}
\usepackage{mathtools}
\usepackage{framed}
\usepackage{epsfig,graphicx,amssymb,amsmath}
\usepackage{diagbox}
\usepackage{color}
\usepackage{setspace}

\usepackage{amsthm}
\usepackage{algorithm}
\usepackage[noend]{algpseudocode}
\usepackage{booktabs}
\usepackage{diagbox}
\usepackage{float}
\usepackage{epstopdf}
\usepackage{ragged2e}
\usepackage{stfloats}
\renewcommand{\raggedright}{\leftskip=0pt \rightskip=0pt plus 0cm}
\usepackage{graphicx}
\usepackage{subfig}  
\usepackage[colorlinks=true,
            linkcolor=blue,
            citecolor=blue,
            urlcolor=blue]{hyperref}
\usepackage{xcolor}
\usepackage[T1]{fontenc}

\ifCLASSINFOpdf

\else

\fi

\usepackage{xcolor}

\usepackage{xpatch}

\begin{document}

\title{DHEA-MECD: An Embodied Intelligence-Powered DRL Algorithm for AUV Tracking in Underwater Environments with High-Dimensional Features}

\author{Kai Tian,
Chuan Lin,~\IEEEmembership{Member,~IEEE},
Guangjie Han,~\IEEEmembership{Fellow,~IEEE},
Chen An,
Qian Zhu,
Shengchao Zhu,\\
Zhenyu Wang

\thanks{\emph{Corresponding author: Guangjie Han}} 
\thanks{Kai Tian, Chuan Lin, Qian Zhu, Zhenyu Wang, are with Software College, Northeastern University, Shenyang, China. (e-mails: kaitian901@gmail.com; chuanlin1988@gmail.com; zhuq@swc.neu.edu.cn; larrywang1019@outlook.com).}
\thanks{Guangjie Han and Shengchao Zhu are with Key Laboratory of Maritime Intelligent Network Information Technology, Ministry of Education, Hohai University. (e-mails: hanguangjie@gmail.com; zhushengchao77@gmail.com.)}
\thanks{Chen An is with Computer Science and Engineering College, Northeastern University, Shenyang, China. (e-mails: 20235854@stu.neu.edu.cn).}
}

\markboth{Journal of \LaTeX\ Class Files,~Vol.~XX, No.~X, XXX~XXXX}%
{Shell \MakeLowercase{\textit{et al.}}: Bare Demo of IEEEtran.cls for IEEE Journals}
%

\IEEEtitleabstractindextext{%
	\begin{abstract}

In recent years, autonomous underwater vehicle (AUV) systems have demonstrated significant potential in complex marine exploration.
However, effective AUV-based tracking remains challenging in realistic underwater environments characterized by high-dimensional features, including coupled kinematic states, spatial constraints, time-varying environmental disturbances, etc.
To address these challenges, this paper proposes a hierarchical embodied-intelligence (EI) architecture for underwater multi-target tracking with AUVs in complex underwater environments.
Built upon this architecture, we introduce the Double-Head Encoder-Attention-based Multi-Expert Collaborative Decision (DHEA-MECD), a novel Deep Reinforcement Learning (DRL) algorithm designed to support efficient and robust multi-target tracking.
Specifically, in DHEA-MECD, a Double-Head Encoder–Attention-based information extraction framework is designed to semantically decompose raw sensory observations and explicitly model complex dependencies among heterogeneous  features, including spatial configurations, kinematic states, structural constraints, and stochastic perturbations.
On this basis, a motion-stage-aware multi-expert collaborative decision mechanism with Top-\textbf{$k$} expert selection strategy is introduced to support stage-adaptive decision-making.
Furthermore, we propose the DHEA-MECD-based underwater multi-target tracking algorithm to enable AUV smart, stable, and anti-interference multi-target tracking.
Extensive experimental results demonstrate that the proposed approach achieves superior tracking success rates, faster convergence, and improved motion optimality compared with mainstream DRL-based methods, particularly in complex and disturbance-rich marine environments.

		
\end{abstract}
	}
\maketitle
\IEEEdisplaynontitleabstractindextext
\IEEEpeerreviewmaketitle

\IEEEpeerreviewmaketitle

\section{Introduction}

\label{sec:introduction}

\IEEEPARstart{A}{quatic} resources are fundamental to human survival and societal development, rendering their sustainable exploration and protection a critical global concern~\cite{oceanexplor}. With the rapid advancement of underwater communication, sensing, and robotic technologies~\cite{underwaterrobots},~\cite{underwaterrobots2}, AUVs have emerged as the primary mobile computing and sensing agents for various maritime missions, ranging from environmental surveillance to high-stakes resource reconnaissance~\cite{auvdevelop}. In such missions, the ability to continuously monitor dynamic entities—such as biological clusters or drifting anomalies—is essential. Thus, autonomous multi-target tracking  stands as a cornerstone capability~\cite{multi_agent},~\cite{tracking2}, requiring the AUV to maintain persistent surveillance of dynamic targets in  underwater spaces. 

Fundamentally, such autonomous tracking necessitates the tight coupling of continuous environmental perception and precise agent control~\cite{autotracking},~\cite{control}.
In this context, EI has risen as a promising emerging paradigm for solving complex agent control problems~\cite{EIcontrol}, integrating perception and execution to withstand underwater disturbances.
Within the paradigm of EI, DRL is widely employed as a core decision-making framework for autonomous agents~\cite{EIDRL}. 
However, conventional DRL algorithms often struggle with robustness and adaptability when applied to complex marine environments, which are filled with unpredictable disturbances and noisy sensory data. Consequently, precise multi-target tracking remains a challenge due to two main technical bottlenecks:

\textbf{1) High-Dimensional Feature Heterogeneity and Perception Fragmentation}: 
Underwater observations are inherently high-dimensional and heterogeneous~\cite{underwatermultisource}, encompassing 3D spatial geometry, kinematic states, and complex environmental variables such as spatiotemporal ocean currents and multi-source acoustic noise.
Traditional DRL algorithms often process these inputs as monolithic flat vectors~\cite{TroDRL}, failing to resolve the intrinsic semantic correlations among sub-features. 
This lack of structural inductive bias leads to ``perception fragmentation", where the agent struggles to distil task-relevant signals from redundant and noisy high-dimensional data~\cite{underwaternoise}, ultimately resulting in policy instability or convergence failure.

\textbf{2) Decision Inadaptability in Dynamic Multi-Stage Tasks}: 
The underwater domain is characterized by intense non-stationarity, requiring AUVs to switch frequently between diverse motion regimes—such as rapid pursuit, precision station-keeping, and emergency collision avoidance~\cite{tracking1}, ~\cite{robotcontrol}. 
Conventional DRL architectures usually rely on a single policy network with fixed parameters. However, a monolithic policy often fails to balance the explosiveness of high-speed maneuvers with the precision of fine-grained control~\cite{roboticmanipulators}, particularly when dealing with the hybrid action spaces (discrete actions coupled with continuous actions)~\cite{hybridaction} required for adaptive underwater tracking.

Inspired by the transformative success of DRL-driven EI architecture in complex robotic control~\cite{RLEI}, this paper proposes a hierarchical EI architecture structured into three integrated layers: the embodied perception layer, the embodied decision layer, and the embodied execution layer. 
Diverging from conventional monolithic DRL, this architecture decouples high-level strategic tracking logic from low-level hydrodynamic actuation. 
Built upon this EI architecture, this paper proposes the \textbf{Double-Head Encoder-Attention-based Multi-Expert Collaborative Decision (DHEA-MECD)} algorithm to serve as the cognitive core for autonomous underwater tracking. 
The proposed DHEA-MECD algorithm integrates a double-head encoder-attention mechanism with a multi-expert collaborative decision mechanism to ensure robust and precise tracking.
Specifically, it employs specialized multi-head encoders and self-attention mechanisms to semantically decompose and fuse heterogeneous state subspaces, facilitating robust representation learning from high-dimensional features.
On this basis, a motion-stage-aware decision mechanism, driven by a Top-$k$ expert selection strategy, dynamically switch between specialized experts, ensuring precise decision across diverse tracking stages.
This algorithm significantly enhances maneuverability and stability while minimizing computational overhead in underwater environments with high-dimensional features.

The main contributions of this work are summarized as follows:

\begin{itemize}
    \item We propose a hierarchical EI architecture for AUVs, which redefines the perception-to-decision pipeline by decoupling the task (e.g., multi-target tracking) into specialised layers, thereby enhancing system robustness in underwater environments with high-dimensional features.

    \item  We propose DHEA-MECD algorithm, which combines double-head encoder-attention and multi-expert collaborative decision mechanisms, based on the proposed EI architecture. It employs specially designed encoders and self-attention mechanism to distill task-essential features, top-$k$ expert selection strategy achieving superior efficiency and precision.
    
    \item  We propose a DHEA-MECD-based multi-target tracking algorithm designed for AUVs in underwater environments with high-dimensional features. The evaluations demonstrate that our proposed algorithm ensure smooth, robust and precise tracking, outperforming other mainstream algorithms.
\end{itemize}

    The rest of this paper is organized as follows: Section~\ref{Section:2} surveys the related work; Section~\ref{Section:3} presents the preliminaries; Section \ref{Section:4} describes the proposed hierarchical EI architecture; Section~\ref{Section:5} introduces the proposed DHEA-MECD algorithm; Section~\ref{Section:6} proposes a multi-target tracking algorithm based on DHEA-MECD;
    Section~\ref{Section:7} provides the evaluation results; Section~\ref{Section:8} concludes the paper.

\section{Related Works}\label{Section:2}
This section reviews recent advances in related fields, focusing on three key areas: 1) AUV-based underwater target tracking, 2) DRL-based decision-making for AUV, and 3) EI-based decision-making for robots.

\subsection{ AUV-Based Underwater Target Tracking}\label{Section:2-1}
AUVs play a vital role in underwater target tracking. To operate reliably in complex hydrodynamic environments while maintaining precise proximity to moving targets, AUVs demand robust control strategies.
In~\cite{ferri2018autonomous}, a data-driven receding horizon control strategy is proposed to optimize AUV heading for target tracking within sonar networks. By integrating an acoustic perception module, the controller autonomously adjusts the AUV's trajectory to maximize the signal-to-noise ratio, significantly reducing target estimation errors and enhancing tracking stability in noisy environments. 
In~\cite{zhu2024underwater}, the authors introduce a software-defined control architecture to improve the adaptive tracking capabilities of AUVs. This framework decouples control logic from vehicle execution, allowing the AUV to dynamically refine its tracking policy through an incremental learning mechanism. This approach ensures continuous and stable target surveillance while maintaining high operational flexibility without requiring model retraining.
In~\cite{chen2021path}, a path planning method based on behavioral decision-making (PPM-BBD) is developed to optimize the AUV’s trajectory during the diving and approach phase. By incorporating an angle modification strategy into the L-SHADE evolutionary algorithm, the method ensures that all generated tracking paths strictly adhere to the AUV's physical maneuverability limits. This control-aware planning achieves over 9\% energy savings, facilitating efficient, long-duration tracking missions.

Overall, the aforementioned traditional control-based strategies provide efficient manners for scheduling the AUVs in ideal environment. 
However, they often struggle to generalize across highly stochastic underwater environment. To address these adaptive challenges, researchers have increasingly turned to DRL for end-to-end decision-making.

\subsection{DRL-Based Decision-Making for AUVs}\label{Section:2-2}

DRL has emerged as a powerful tool for AUV decision-making, enabling autonomous vehicles to learn optimal control policies in highly nonlinear and uncertain underwater environments.
In~\cite{liang2024underwater}, an ACL-SAC algorithm is proposed to optimize AUV yaw and propulsion force for dynamic tracking. By integrating an Attention Conv-LSTM (ACL) with Soft Actor-Critic (SAC), the framework enhances the AUV's robustness against ocean currents and acoustic uncertainties. This DRL approach ensures rapid convergence and superior adaptability to stochastic environmental disturbances compared to conventional methods.
In~\cite{fang2022auv}, a fast-deployed DDPG-based controller is developed for the posture control and trajectory tracking of under-actuated X-rudder AUVs. The method effectively resolves control failures at critical yaw angles, achieving precise three-degree-of-freedom tracking control (pitch, yaw, and speed). This DRL-driven strategy facilitates rapid deployment and stable position tracking for targets in various orientations.
In~\cite{zhu2024underwater2}, an Interrupted Software-Defined DRL architecture (ISD-MARL) is introduced to accelerate AUV decision-making. By employing the S-MADDPG algorithm with an action optimization model, the framework significantly shortens policy convergence time. Combined with an interrupted path planning scheme, it enables the AUV to maintain a balance between tracking precision and time efficiency in dynamic underwater scenarios.

In summary, DRL significantly improves the AUV's response to environmental uncertainties. 
However, most DRL models lack a deep understanding of physical interaction and environmental reasoning. This limitation has motivated the exploration of EI to better synergize physical form with cognitive processes.

\subsection{EI-Based Decision-Making for Robots}~\label{Section:2-3}
EI shifts the robotic paradigm from pure data-driven learning to physical interaction, enabling robots to perform reasoning and decision-making within structured and unstructured environments.
In~\cite{gan2024embodied}, a bionic robot controller is proposed that integrates environment perception, autonomous planning, and motion control into a unified framework. By deploying the YOLOv5\_OBB network for object recognition and an improved RRT-Growth-Angle algorithm for trajectory planning, the controller mimics human-like decision-making for complex tasks such as multi-object rearrangement. This modular integration ensures precise dual-robot coordination and collision-free execution in customized manufacturing scenarios.
In~\cite{long2023human}, the authors investigate human-in-the-loop EI by developing an interactive simulation platform for surgical robot learning. The study emphasizes the role of human demonstrations in refining policy learning through reinforcement learning. By facilitating high-quality human-robot interaction via physical input devices, the framework enables surgical robots to learn complex control policies more efficiently, demonstrating the potential of EI in safety-critical and high-precision domains.
In~\cite{song2025embodied}, this research highlights how EI redefines robotic manipulation by bridging the gap between digital data and physical action. Through improved reasoning and generalization, EI enables robots to navigate the complexities of unstructured settings. Ultimately, the authors identify EI as a fundamental building block for AGI, providing the theoretical grounding necessary for deploying intelligent, autonomous systems in the field.These advances illustrate how EI bridges digital reasoning with physical action. However, existing EI frameworks often incur prohibitive training costs and neglect the specific constraints of underwater mobility. To bridge this gap, we propose a hierarchical EI architecture for underwater smart tracking, which enhances both efficiency and adaptability in underwater environments with high-dimensional features.

\section{Preliminaries}\label{Section:3}

In this section, we introduce the preliminary knowledge related to this work, including ocean current disturbance modeling, underwater noise and the Markov Decision Process (MDP) for AUV-based multi-target tracking.

\subsection{Ocean Current Disturbance Modeling}\label{3-1}
To characterize the spatiotemporal variability of ambient flow, the ocean current velocity field is represented as a finite-dimensional, linearly parameterized model based on Gaussian radial basis functions (RBFs). Each basis function $\phi_\ell(r)$ is defined as:
\begin{equation}
\phi_\ell(r) = \exp\!\left( -\frac{\|r-\mu_\ell\|^2}{2\lambda_\ell^2} \right), \quad \ell = 1,\ldots,L,
\end{equation}
where $r \in \mathbb{R}^3$ denotes the spatial coordinate, while $\mu_\ell \in \mathbb{R}^3$ and $\lambda_\ell > 0$ represent the center and spatial scale of the $\ell$-th basis function, respectively.

The ocean current velocity ${v}_c(r,t)$ at location $r$ and time $t$ is expressed as:
\begin{equation}
\ v_c(r,t) = \Phi(r)\,\theta(t),
\label{eq:current_linear_param}
\end{equation}
where $\theta(t) \in R^{3L}$ is a vector of time-varying weights, and the regression matrix $\Phi(r) = [\phi_1(r)·I_3, \ldots, \phi_L(r)·I_3] \in R^{3 \times 3L}$ is constructed from the RBFs, with $I_3$ denoting the $3 \times 3$ identity matrix.

The ambient flow induces relative motion between the AUV and the surrounding fluid. Let $p_a(t) \in R^3$ denote the position of the AUV in the inertial frame, and $v_a(t) = \dot{p}_a(t)$ be its corresponding inertial velocity. The relative velocity $v_r(t)$ with respect to the ambient flow is then defined as:
\begin{equation}
v_r(t) = v_a(t) - \Phi\bigl(p_a(t)\bigr)\,\theta(t),
\label{eq:relative_velocity_param}
\end{equation}
where $\Phi\bigl(p_a(t)\bigr)\,\theta(t)$ represents the local current velocity evaluated at the AUV's current position.

Accordingly, the translational dynamics of the AUV satisfy the following:
\begin{equation}
M\dot{v}_a(t) = f_u(t) + f_h\bigl(v_r(t)\bigr) + f_d(t),
\label{eq:auv_dynamics}
\end{equation}
where $M \in R^{3 \times 3}$ is the inertia matrix, $f_u(t)$ denotes the control input, $f_h(\cdot)$ represents the hydrodynamic forces dependent on the relative velocity, and $f_d(t)$ captures unmodeled dynamics and external disturbances.

\subsection{Underwater Noise Modeling}\label{3-2}
The instantaneous underwater acoustic noise \(n(t)\) results from the confluence of multiple concurrent sources, including vehicle-induced noise \(n_{\mathrm{veh}}(t)\), biological noise \(n_{\mathrm{bio}}(t)\), geological noise \(n_{\mathrm{geo}}(t)\), and turbulence-induced noise \(n_{\mathrm{turb}}(t)\). Totally, \(n(t)\) can be given by the following:

\begin{equation}\label{eq:noise_superposition_refined}
    n(t) = \big[ n_{\mathrm{veh}}(t),\; n_{\mathrm{bio}}(t),\; n_{\mathrm{geo}}(t),\; n_{\mathrm{turb}}(t) \big]^\top.
\end{equation}

\textbf{Vehicle-induced noise \(n_{\mathrm{veh}}(t)\)} arises from mechanical vibrations and hydrodynamic interactions inherent to underwater platforms. These phenomena typically manifest as broadband fluctuations~\cite{noise1}, with the power spectral density scaling alongside operational intensity. For algorithmic evaluation, we model the \(n_{\mathrm{veh}}(t)\) as a zero mean Gaussian process:

\begin{equation}
n_{\mathrm{veh}}(t)\sim\mathcal{N}\!\left(0,\sigma_{\mathrm{veh}}^{2}\right),
\end{equation}
where the variance $\sigma_{\mathrm{veh}}^{2}$ is given by
\begin{equation}
\sigma_{\mathrm{veh}}^{2}
=\sigma_{\mathrm{veh},0}^{2}
\left(\frac{v_s}{v_0}\right)^{\alpha_v}
\left(\frac{\rho_N}{\rho_0}\right)^{\alpha_\rho},
\end{equation}
where $\sigma_{\mathrm{veh},0}^{2}$, $v_0$, and $\rho_0$ are reference values and $\alpha_v,\alpha_\rho>0$ control sensitivity to speed and density.

\textbf{Biological noise \(n_{\mathrm{bio}}(t)\)} originates from marine fauna vocalizations and impulsive acoustic events, typically exhibiting heavy-tailed, non-Gaussian characteristics. To capture this impulsiveness, we employ a symmetric $\alpha$-stable ($S\alpha S$) distribution~\cite{SaS}:
\begin{equation}
n_{\mathrm{bio}}(t) \sim S\alpha S(\alpha, \gamma, 0),
\end{equation}
where $\alpha \in (1,2]$ and $\gamma > 0$ denote the impulsiveness and scale parameters, respectively. 

\textbf{Geological noise \(n_{\mathrm{geo}}(t)\)} is dominated by low frequency content and exhibits pronounced temporal correlation. 
Thus, we model \(n_{\mathrm{geo}}(t)\) as colored Gaussian noise obtained by filtering white Gaussian excitation:

\begin{equation}\label{eq:geo_filtered}
\left\{
\begin{array}{ll}
n_{\mathrm{geo}}(t) = (h_{\mathrm{geo}} \circledast w)(t); \\
w(t) \sim \mathcal{N}(0,1),
\end{array}
\right.
\end{equation}
where \(h_{\mathrm{geo}}\) is a shaping filter and \(\circledast\) denotes convolution. To enforce low-frequency dominance, \(h_{\mathrm{geo}}\) is selected such that the resulting power spectral density follows a power-law profile:
\begin{equation}
|H_{\mathrm{geo}}(\omega)|^2 \propto \omega^{-\beta_{\mathrm{geo}}},
\end{equation}
where \(H_{\mathrm{geo}}(\omega)\) is the frequency response of \(h_{\mathrm{geo}}\) and \(\beta_{\mathrm{geo}}\) governs the spectral roll-off.

\textbf{Turbulence-induced noise \(n_{turb}(t)\)} arises from flow-driven pressure and vortex fluctuations on sensor nodes and structures, causing temporally correlated amplitude variations~\cite{noise2}. At discrete time index \(t\), we model \(n_{turb}(t)\) as a stochastic process with time-varying intensity:

\begin{equation}\label{eq:turb_model}
\begin{cases}
n_{\mathrm{turb}}(t) = \rho\, n_{\mathrm{turb}}(t-1)
+ \sigma_{\mathrm{turb}}\bigl(U_{\mathrm{turb}}(t)\bigr)\, \epsilon(t); \\[6pt]
\epsilon(t) \overset{\mathrm{}}{\sim} \mathcal{N}(0,1); \\[6pt]
\sigma_{\mathrm{turb}}\bigl(U_{\mathrm{turb}}(t)\bigr) 
= \sigma_{\mathrm{turb},0} \left(\frac{U_{\mathrm{turb}}(t)}{U_0}\right)^{\alpha_u},
\end{cases}
\end{equation}
where \(|\rho| < 1\) controls temporal correlation, \(U_{\mathrm{turb}}(t)\) denotes the (possibly time-varying) turbulence speed or an equivalent intensity indicator, and $\sigma_{\mathrm{turb},0}$, $U_0$, $\alpha_u > 0$ are reference and sensitivity parameters. 
In particular, $U_{\mathrm{turb}}(t)$ is treated as slowly varying within each processing window to yield a locally stationary model.

\subsection{MDP  for Multi-Target Tracking}\label{3-3}

In this paper, the AUV-based multi-target tracking problem in  underwater environments with high-dimensional features is modeled as an MDP, formally defined as
$\mathcal{M} = (\mathcal{S}, \mathcal{A}, P, R, \gamma)$, where $\mathcal{S}$ is the state space, $\mathcal{A}$ is the hybrid action space, $P$ represents the state transition probability, $R$ denotes the reward function, and $\gamma \in [0, 1)$ is the discount factor.

\subsubsection*{State Space $S$}
The state captures all necessary information required for the AUV's decision-making process, including AUV's space information (position, velocity, and attitude), target information (target position, inter-target distances), barrier information (ocean currents, obstacle states), and underwater noise (vehicle-induced noise, biological noise, geological noise and turbulence-induced noise) 

\subsubsection*{Action Space $\mathcal{A}$}
The AUV tracking is realized over a hybrid action space $\mathcal{A} \triangleq \mathcal{D} \times \mathcal{C}$, where each composite action $a_t = (d_t, c_t)$ couples a discrete action $d_t \in \mathcal{D}$ with a continuous action $c_t \in \mathcal{C}$. Here, $d_t$ selects the discrete action (e.g., acceleration, yaw, turn, or braking), while $c_t = [\dot{v}_t, \delta_{\psi,t}, \delta_{\theta,t}, b_t]^\top$ refines the selected discrete action with precise control magnitudes, encompassing magnitude of acceleration, attitude rates, and braking strength. 

\subsubsection*{ State Transition Probability $P$}
State transition dynamics are described by the transition probability $P(s_{t+1} | s_t, a_t)$, which characterizes the stochastic evolution of the AUV-target state. 

\subsubsection*{Reward Function $R$}
The reward function \(R\) is designed to ensure accurate target tracking, and the details will be introduced in Section~\ref{Section:6-1}.

\subsubsection*{Discount Factor $\gamma$}
The discount factor $\gamma \in (0,1)$ balances immediate and long-term performance, enabling the AUV to achieve stable and robust tracking behaviors over tracking missions.

\section{Hierarchical Embodied-Intelligence Architecture for Underwater Smart Tracking}\label{Section:4}

\begin{figure*}[bth]
	\centering
	\includegraphics[width=1.0\linewidth]{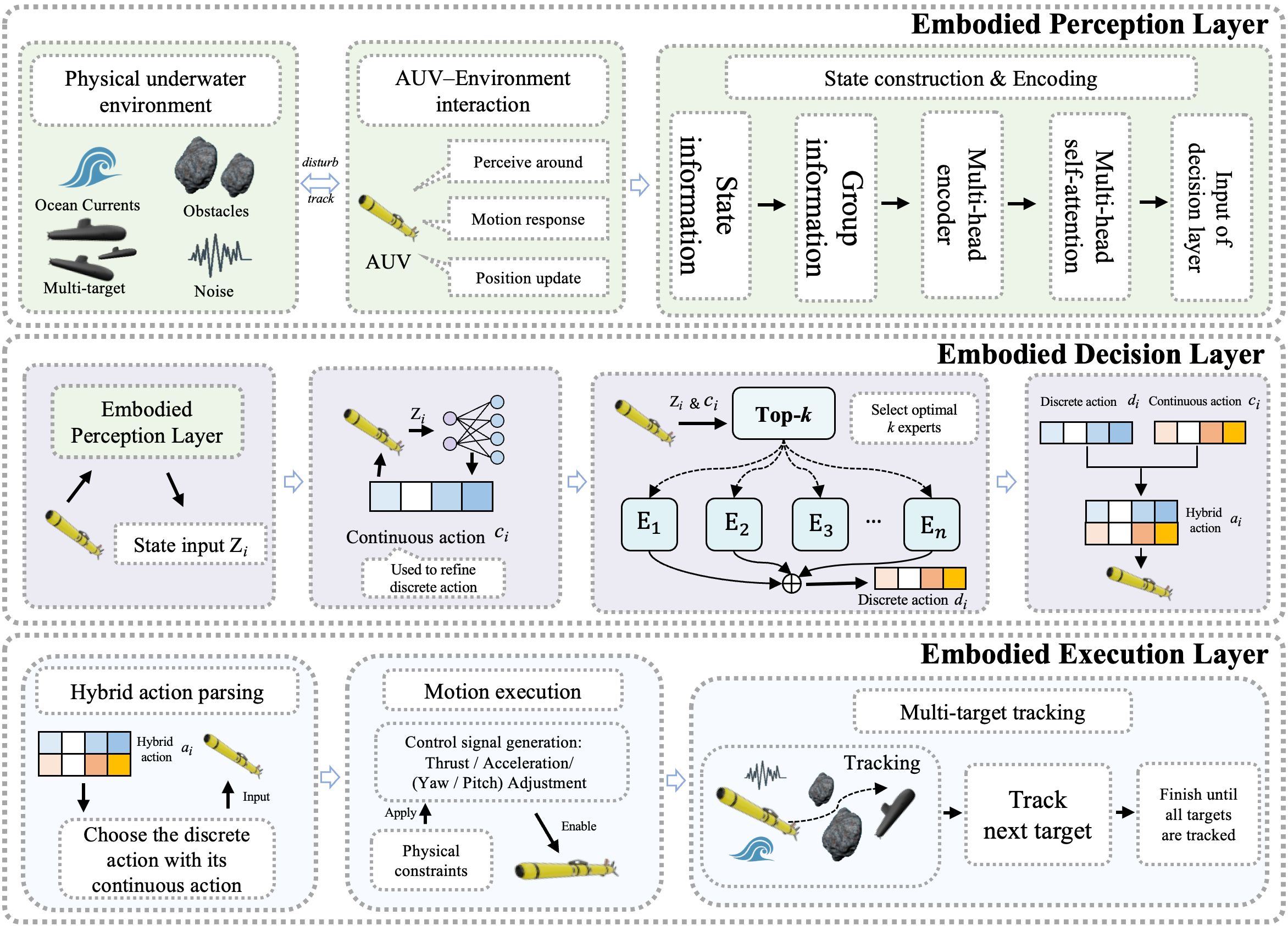}
	\caption{{Proposed hierarchical EI architecture}}
	\label{fig1}
\end{figure*}

In this paper, we propose a hierarchical EI architecture (as shown in Fig.~\ref{fig1}) to facilitate precise and efficient decision-making in multi-target tracking across dynamic underwater environments.
This architecture empowers the AUV with sustained surveillance and precise proximity capabilities by synergizing real-time perception with adaptive decision, even amidst high-dimensional uncertainties and severe environmental disturbances.

The proposed architecture includes an embodied perception layer, an embodied decision layer, and an embodied execution layer:

\textbf{Embodied Perception Layer:} 
To address the challenges of perception in complex underwater environments, this layer is designed to construct a robust and high-fidelity environmental state representation. 
Recognizing that multi-source sensory information possesses distinct physical properties and statistical distributions, the architecture decouples the raw state vector into four specialized subspaces: space, motion, barrier, and noise. Each subspace is processed by a dedicated encoder specifically tailored to its unique characteristics, ensuring that domain-specific features are distilled without mutual interference.
The primary objective of this design is to synthesize these heterogeneous sub-features into a state representation.
By utilizing a multi-head self-attention mechanism, the layer actively models the cross-modal correlations between disparate information streams.
This process yields a unified and context-aware state representation, providing a high-fidelity foundation to empower robust and adaptive decision-making in subsequent stages.

\textbf{Embodied Decision Layer:}
Based on the processed information from the embodied perception layer, the embodied decision layer adopts a multi-expert collaborative decision mechanism (detailed in Section \ref{Section:5-2}), serving as the core policy for underwater applications.
It translates perceptual inputs into executable actions by producing a hybrid action that integrates discrete action with continuous action. 
Leveraging a multi-expert collaborative decision mechanism with the Top-$k$ expert selection strategy, this layer ensures computational efficiency, policy adaptability, and high decision accuracy, enabling the AUV to perform complex tasks in underwater environments with high-dimensional features.

\textbf{Embodied Execution Layer:}
The execution layer serves as the physical interface that bridges high-level strategic intent with the AUV's underwater dynamics. It parses the hybrid action $a_t$  into discrete-continuous control commands specifically tailored for the vehicle's propulsion system. By decoupling qualitative maneuvers (e.g., acceleration action) from quantitative control magnitudes (e.g., magnitude of acceleration), this layer ensures the precise mapping of policy outputs onto physical actuation.

\section{Double-Head Encoder-Attention-based Multi-Expert Collaborative Decision Algorithm}\label{Section:5}

Building upon the hierarchical EI architecture, this paper introduces a novel DRL-based approach: the \textbf{D}ouble-\textbf{H}ead \textbf{E}ncoder-\textbf{A}ttention-based \textbf{M}ulti-\textbf{E}xpert \textbf{C}ollaborative \textbf{D}ecision (DHEA-MECD) algorithm for AUV precise multi-target tracking in underwater environments with high-dimensional features. 
Inspired by the Transformer architecture, the proposed method synergizes a multi-head encoder and multi-head self-attention module with a multi-expert collaborative decision mechanism.
This design leverages a multi-head encoder in conjunction with a self-attention mechanism to facilitate the deep fusion of heterogeneous environmental features from AUV's perceived state space. To translate these representations into precise actions, a Top-$k$ expert selection strategy is introduced, which dynamically selects the most relevant experts based on the real-time motion stage. 
This selection strategy enables motion-stage-aware collaborative decision-making, significantly enhancing tracking accuracy and computational efficiency in underwater environments with high-dimensional features.

\begin{figure*}[bth]
	\centering
	\includegraphics[width=1\linewidth]{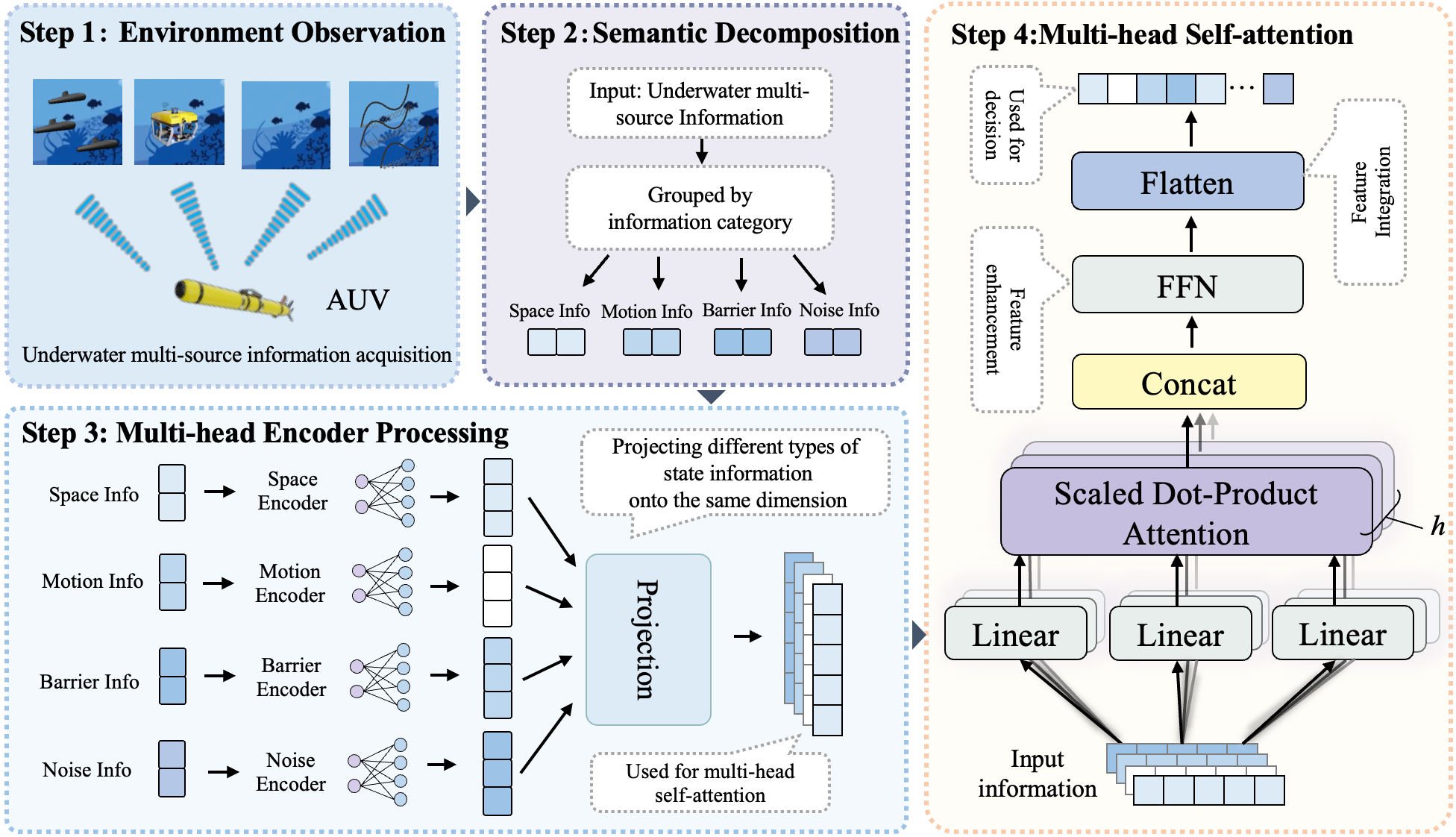}
	\caption{{Proposed information extraction framework for high-dimensional features in DHEA-MECD}}
	\label{fig3}
\end{figure*}

\subsection{Double-Head Encoder-Attention-Driven Information Extraction}\label{Section:5-1} Accurate environmental perception is the cornerstone of intelligent decision-making, enabling AUVs to execute precise tracking in underwater environments with high-dimensional features. Conventional methods typically input raw state vectors directly into neural networks, which frequently fails to capture the intrinsic correlations among heterogeneous semantic features. 
To facilitate robust decision-making in such complex scenarios, we propose a structured state representation and design a multi-head encoder architecture designed to extract meaningful latent features from sensory observations. As illustrated in Fig. \ref{fig3}, the integrated processing pipeline is composed of four sequential steps:

\textbf{Step 1 - Environment Observation:}
At each decision step, the AUV acquires a high-dimensional observation vector $s$, which encapsulates multi-source information from the underwater environment. This vector includes spatial geometry, motion dynamics, constraint conditions, and stochastic disturbances arising from sensing uncertainties and external perturbations.

\textbf{Step 2 - Semantic Decomposition:}

To ensure numerical consistency, each component of the observation vector $s$ is first normalized to the range $[0,1]$. The state vector is decomposed into four semantically independent sub-vectors: $s = [s_{\text{sp}}; s_{\text{mot}}; s_{\text{bar}}; s_{\text{noise}}]^\top$. 
First, $s_{\text{sp}}$ denotes the space information, including the AUV's position $p_a(t)$, the positions of all targets $j \in \mathcal{J}$ (denoted as $p_j(t)$), the Euclidean distances $d_{a,j}(t)$, and a capture indicator $I(d_{a,j}(t) < r_{\text{target}})$. 
Second, $s_{\text{mot}}$ describes the motion state, comprising the translational velocity $v_a(t)$, sinusoidal encodings of yaw-pitch angles $(\psi_a(t), \theta_a(t))$, and a normalized time index $t_{\mathrm{ratio}} = t / T_{\max}$. 
Third, $s_{\text{bar}}$ represents barrier information, which encapsulates external constraints hindering AUV tracking. This component includes the local ocean current velocity obstruction $v_c(p_a(t), t)$ and a set of $N$ surrounding obstacles, denoted by $p_{o,i}(t)$ for $i = 1, \dots, N$.
Finally, $s_{\text{noise}}$ is the noise information which captures stochastic disturbances affecting sensing and actuation.
Consistent with the underwater noise model in Section~\ref{3-2}, it is represented by a noise state vector
$n(t) = \bigl[n_{\mathrm{veh}}(t),\, n_{\mathrm{bio}}(t),\, n_{\mathrm{geo}}(t),\, n_{\mathrm{turb}}(t)\bigr]^\top$,
which aggregates multiple noise mechanisms into a structured stochastic input.

To address the heterogeneous physical properties  in these sub-vectors, they are inputted into four specialized encoders. This modular design ensures that domain-specific features are distilled through targeted extraction before being integrated into the global policy representation.

\textbf{Step 3 - Multi-Head Encoder Processing:}
To further process the information, we design a multi-head encoder composed of four specialized sub-encoders. Each module performs specific feature extraction and generates an aligned latent representation.

\textbf{Space encoder} employs a feedforward network with GELU~\cite{gelu} activation and layer normalization:
\begin{equation}
\left\{
\begin{aligned}
    \widetilde{h}_s^{(l)} &= \mathrm{GELU}(W_s^{(l)} h_s^{(l-1)} + b_s^{(l)}); \\
    h_s^{(l)} &= \mathrm{LayerNorm}(h_s^{(l-1)} + \widetilde{h}_s^{(l)}),
\end{aligned}
\right.
\end{equation}
where $h_s^{(l)} \in \mathbb{R}^{d}$ denotes the spatial representation at layer $l$,
$W_s^{(l)} \in \mathbb{R}^{d \times d}$ and $b_s^{(l)} \in \mathbb{R}^{d}$ are the learnable parameters, 
and $\widetilde{h}_s^{(l)}$ is the post-activation intermediate output. 
Note that LayerNorm normalizes the feature dimension of each sample and stabilizes the residual propagation.

\textbf{Motion encoder} is for extracting the motion information.
In our approach, the motion features are processed using a Swish-activated MLP followed by GroupNorm~\cite{gn}:
\begin{equation}
\left\{
\begin{aligned}
    \widetilde{h}_m^{(l)} &= \mathrm{Swish}(W_m^{(l)} h_m^{(l-1)} + b_m^{(l)}); \\
    h_m^{(l)} &= \mathrm{GroupNorm}(\widetilde{h}_m^{(l)}),
\end{aligned}
\right.
\end{equation}
where $h_m^{(l)} \in \mathbb{R}^{m}$ is the motion feature vector at layer~$l$,  
$W_m^{(l)}$ and $b_m^{(l)}$ are trainable parameters,  
and GroupNorm normalizes features across pre-defined channel groups, making the representation insensitive to batch size and suitable for motion signals with varying magnitudes.

\textbf{Barrier encoder} is built based on a ReLU-based residual MLP:
\begin{equation}
\left\{
\begin{aligned}
    \widetilde{h}_b^{(l)}
        &= \mathrm{ReLU}\!\left(W_b^{(l)} h_b^{(l-1)} + b_b^{(l)}\right); \\
    h_b^{(l)}
        &= \mathrm{LayerNorm}\!\left(h_b^{(l-1)} + \widetilde{h}_b^{(l)}\right),
\end{aligned}
\right.
\end{equation}
where $h_b^{(l)} \in \mathbb{R}^{d}$ represents the barrier features,  
$W_b^{(l)}$ and $b_b^{(l)}$ are the corresponding trainable parameters,  
and $\widetilde{h}_b^{(l)}$ is the activated intermediate output.  
It should be clarified that the residual pathway ensures stable gradient flow for the often high-variance constraint-related features.

\textbf{Noise encoder} is dedicated to extracting the meaningful information from the noise by using a ReLU-activated MLP followed by Batch Normalization:
\begin{equation}
\left\{
\begin{aligned}
    \widetilde{h}_n^{(l)}
        &= \mathrm{ReLU}\!\left(W_n^{(l)} h_n^{(l-1)} + b_n^{(l)}\right); \\
    h_n^{(l)}
        &= \mathrm{BatchNorm}\!\left(\widetilde{h}_n^{(l)}\right),
\end{aligned}
\right.
\end{equation}
where vector $h_n^{(l)} \in \mathbb{R}^{d}$ represents the noise information,
$W_n^{(l)}$ and $b_n^{(l)}$ are learnable parameters,  
and BatchNorm normalizes activations using batch-wise statistics, effectively suppressing stochastic fluctuations in noisy measurements.

After the final layer of each encoder, the modality-specific representation $h_i \in \mathbb{R}^{d}$  
($i \in \{s, d, e, n\}$) is mapped to a shared latent space $t_i$:
\begin{align}
    t_i = W_i h_i + b_i, \qquad t_i \in \mathbb{R}^{n},
\end{align}
where $W_i \in \mathbb{R}^{n \times d}$ and $b_i \in \mathbb{R}^{n}$ are projection parameters for each modality.  
The four projected embeddings form a feature sequence
\begin{align}
    T = [t_s ; t_d ; t_e ; t_n] \in \mathbb{R}^{4 \times n},
\end{align}
which is then fed into the multi-head self-attention module.

\textbf{Step 4 - Multi-Head Self-Attention:}
To explicitly model the dependencies among the four specific vectors, we employ a multi-head self-attention module. For each head, the query $Q$, key $K$, and value $V$ matrices are computed as
\begin{align}
Q = T  W_Q;\quad
K = T  W_K;\quad
V = T  W_V,
\end{align}
where $T$ is the input information matrix,  
$W_Q, W_K, W_V$ are the learnable projection matrices for queries, keys, and values, respectively.  
Thus, the head-wise attention output is obtained by:
\begin{align}
A_h = \mathrm{Softmax}\!\left(\frac{Q_h K_h^\top}{\sqrt{d_h}}\right)V_h,
\end{align}
where
$A_h$ is the attention output of the $h$-th head,  
$Q_h, K_h, V_h$ denote the head-specific query, key, and value matrices,  
$d_h$ is the dimensionality of each attention head.
All the head outputs are concatenated and linearly fused:
\begin{align}
M = \mathrm{Concat}(A_1,\ldots,A_6) W_O,
\end{align}
where $A_1,\ldots,A_6$ are outputs of all the six attention heads,  
$\mathrm{Concat}(\cdot)$ denotes the concatenation along the feature dimensional,  
$W_O$ is the output projection matrix.  

Afterwards, a feedforward enhancement block is employed to further refine the fused information representation \(M\):
\begin{equation}\label{eq:ffn_block}
\left\{
\begin{aligned}
z_1 &= \mathrm{GELU}\!\left(W^{f}_1 M + b^{f}_1\right); \\
z_2 &= \mathrm{Dropout}\!\Bigl(\mathrm{GELU}\!\left(W^{f}_2 z_1 + b^{f}_2\right)\Bigr); \\
Z &= \mathrm{LayerNorm}\!\left(W^f_3 z_2 + b^{f}_3\right) +M,
\end{aligned}
\right.
\end{equation}

where \(W^{f}_1, W^{f}_2, W^{f}_3\) and 
\(b^{f}_1, b^{f}_2, b^{f}_3\) are learnable parameters of the feedforward block.
The variables \(z_1\) and \(z_2\) denote intermediate feature transformations.
\(\mathrm{GELU}(\cdot)\) is the Gaussian error linear unit activation, 
\(\mathrm{Dropout}(\cdot)\) is applied for regularization, and 
\(\mathrm{LayerNorm}(\cdot)\) denotes layer normalization.
The residual connection with input \(M\) is introduced to stabilize optimization and facilitate gradient propagation.

The enhanced information sequence \(Z\) is subsequently flattened and projected into a unified policy embedding through a sequence of fully connected layers:
\begin{equation}\label{eq:ffn_block}
\left\{
\begin{aligned}
h_1 &= \mathrm{LayerNorm}\!\left(
\mathrm{GELU}\!\left(W^{p}_1 z_f + b^{p}_1\right)
\right); \\
h_2 &= \mathrm{LayerNorm}\!\Bigl(
\mathrm{Dropout}\!\left(
\mathrm{GELU}\!\left(W^{\mathrm{p}}_2 h_1 + b^{p}_2\right)
\right)
\Bigr),
\end{aligned}
\right.
\end{equation}
where \(z_f\) denotes the flattened version of \(Z\), and
\(W^{p}_1, W^{p}_2\) and \(b^{p}_1, b^{p}_2\) are the parameters of the projection layers.
Finally, the policy embedding is obtained as
\begin{equation}
z = W^{p}_3 h_2 + b^{p}_3,
\end{equation}
where the vector \(z\) encodes cross-structural dependencies among high-dimensional features and provides the decision layer with a stable and coherent environmental representation.

Overall, the proposed multi-head encoder and multi-head self-attention module hierarchically integrate information interactions and global structural dependencies, yielding a compact and stable representation for policy learning.

\subsection{Multi-Expert  Collaborative Decision Mechanism }\label{Section:5-2}

\begin{figure*}[t!]
	\centering
	\includegraphics[width=1\linewidth]{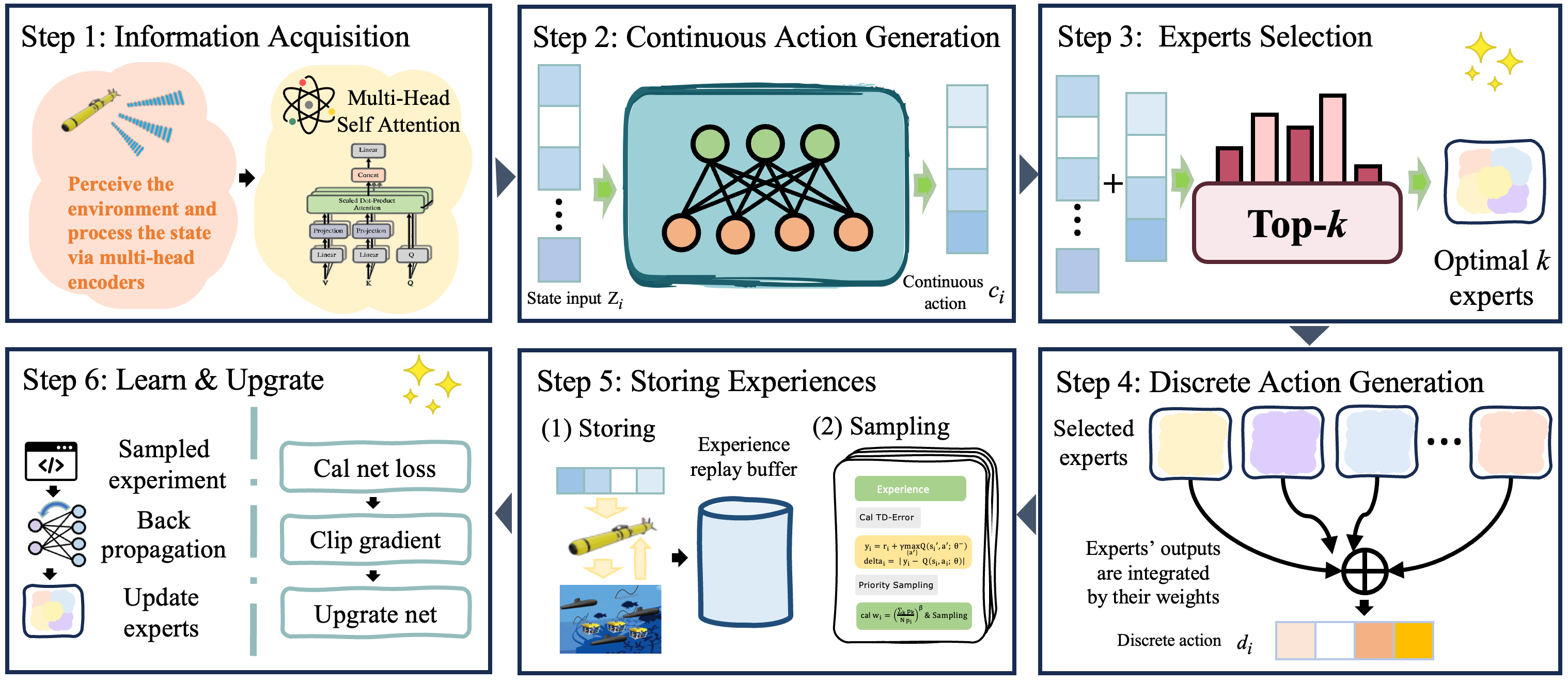}
	\caption{Proposed  Multi-Expert  Collaborative Decision  Mechanism in DHEA-MECD}
	\label{fig4}
\end{figure*}

Built upon multi-modal environmental representation extracted by the proposed information extraction framework, this section presents the proposed multi-expert collaborative decision mechanism which employs a motion-stage-aware Top-$k$ strategy to achieve adaptive tracking in complex underwater environments.
By selectively invoking the $k$ most relevant experts based on real-time motion stages, the mechanism ensures specialized decision-making while minimizing the per step computational burden. Fig.~\ref{fig4} depicts the overall algorithm, followed by a detailed description of its core steps.

\textbf{Step 1 - Information Acquisition:}
Prior to decision-making, the raw multi-source observation $s$ is processed via the information extraction framework detailed in Section \ref{Section:5-1}. This yields a compact latent state vector \(z_t\), which encapsulates fused spatiotemporal dependencies and serves as the foundational input for the subsequent hybrid action generation.

\textbf{Step 2 - Continuous Action Generation:}
To enable fine-grained control beyond discrete actions, we design a network that maps the state \(z_t\) to continuous action parameters, forming a hybrid action space. Structurally, this network employs a GRU to capture temporal dynamics, followed by an MLP for nonlinear feature transformation.

First, the GRU updates the hidden state \(h_t\) to model historical context:
\begin{equation}
    h_t = \mathrm{GRU}(z_t, h_{t-1}),
\end{equation}
where \(h_t\) serves as a compact temporal representation.
Subsequently, \(h_t\) is processed by a two-layer ReLU-activated MLP and projected into the action space via a hyperbolic tangent output layer:
\begin{equation}
\left\{
\begin{aligned}
    f_1 &= \mathrm{ReLU}(W^{c}_1 h_t + b^{c}_1); \\
    f_2 &= \mathrm{ReLU}(W^{c}_2 f_1 + b^{c}_2); \\
    c_t &= \tanh(W^{c}_o f_2 + b^{c}_o),
\end{aligned}
\right.
\label{eq:continuous_policy}
\end{equation}
where \(W^{c}_{(\cdot)}\) and \(b^{c}_{(\cdot)}\) denote the learnable weights and biases for each layer. The \(\tanh\) activation constrains the output to \((-1,1)\), ensuring bounded control.

The resulting continuous action vector \(c_t \in \mathbb{R}^4\) encodes the normalized acceleration \(\dot{v}_t\), yaw adjustment \(\delta_{\psi,t}\), pitch adjustment \(\delta_{\theta,t}\), and braking intensity \(b_t\), effectively refining the AUV's motion execution.

\textbf{Step 3 - Experts Selection:}
To dynamically identify the most relevant experts for the current motion stage, we employ a Top-\(k\) expert selection strategy.
For each expert \(i\), a learnable key vector \(k_i\) is introduced to encode its functional specialization. The matching score \(g_i\) is computed by projecting the state \(z_t\) onto the expert space, followed by a Top-\(k\) selection strategy to determine the selected expert subset \(\Omega_t\):
\begin{equation}
\left\{
\begin{aligned}
    g_i &= z_t^{\top}W_g k_i + b_i; \\
    \Omega_t &= \mathrm{Topk}(\{g_1, \dots, g_n\},\, k).
\end{aligned}
\right.
\label{eq:gate_score}
\end{equation}
Subsequently, a sparse softmax is applied over \(\Omega_t\) to obtain the normalized experts weights:
\begin{equation}
    \alpha_i(z_t) = \frac{\exp(g_i)}{\sum_{j \in \Omega_t} \exp(g_j)}, \quad i \in \Omega_t.
    \label{eq:sparse_softmax}
\end{equation}
This sparse selection ensures computational efficiency while focusing model capacity on the most relevant experts.

Structurally, each expert \(i\) comprises a GRU-MLP backbone (with independent parameters) and a motion-stage-aware perception module. The expert output \(E_i(z_t)\) is formulated as:
\begin{equation}
    E_i(z_t) = \mathrm{MLP}_i(\mathrm{GRU}_i(z_t)) + \underbrace{\sum_{j=1}^{d} w_{ij} f_{ij}(z_t^{(j)})}_{P_i(z_t)} + b^{E}_i,
    \label{eq:expert}
\end{equation}
where the motion-stage-aware perception module \(P_i(z_t)\) applies weighted transformations \(f_{ij}\) to specific state components, enhancing expert-level specialization by selectively emphasizing features critical to the assigned motion stage.

\textbf{Step 4 - Discretized Action Generation:} 
Based on the continuous action  \(c_t\) generated in Step 2, this step identifies the optimal discrete action \(d_t\) at the current time step \(t\). Since \(c_t\) has already refined the granularity of action, the decision policy evaluates each candidate discrete action \(d \in \mathcal{D}\) by pairing it with the current state \(z_t\) and continuous action \(c_t\).

Conditioned on the selected expert subset \(\Omega_t\), the global hybrid action-value for a specific pair \((d, c_t)\) is derived by aggregating the specialized evaluations via the normalized expert weights:
\begin{equation}
    Q_{\text{final}}(z_t, d, c_t; \omega) 
    = \sum_{i \in \Omega_t} \alpha_i(z_t) \cdot Q_i(z_t, d, c_t; \omega_i),
    \label{eq:qfinal}
\end{equation}
where \(i\) denotes the expert index, and \(\omega\) represents the selected expert subset. 
Ultimately, the agent selects the discrete action that yields the highest expected return under the current expert consensus:
\begin{equation}
    d_t = \underset{d \in \mathcal{D}}{\arg\max} \; Q_{\text{final}}(z_t, d, c_t; \omega).
\label{finaldt}
\end{equation}

\textbf{Step 5 - Experience Storage:}
Transitions are encapsulated as tuples \(e_t = (z_t, d_t, c_t, r_t, z_{t+1}, done_t)\) and stored in a Prioritized Experience Replay (PER)~\cite{per} buffer \(\mathcal{B}\). 
To enhance sample efficiency, during sampling, the priority of each transition \(j\) is updated based on the Temporal-Difference (TD) error \(\delta_j\):
\begin{equation}
    \delta_j = \left| y_j - Q_{\text{final}}(z_j, d_j, c_j; \omega) \right|,
    \label{eq:td_error}
\end{equation}
where \(j\) denotes the index of the sampled transition in the minibatch, \(y_j\) is the TD target, and \(c_j\) is the executed continuous action. This strategy biases the sampling process towards transitions with higher learning potential.

\textbf{Step 6 - Learning and Update:} 
The training objective is to jointly optimize the expert network parameters \(\omega\) and the continuous action network parameters \(\theta\). Minibatches of size \(B_{batch}\) are sampled from \(\mathcal{B}\) using importance sampling weights \(w_j\).

\textbf{Experts Network Update:}
The primary objective of the multi-expert network is to minimize the Bellman prediction error. Specifically, the TD target \(y_j\) for transition \(j\) is constructed by maximizing over the discrete action  utilizing the target continuous policy:
\begin{equation}
    y_j = r_j + \gamma (1 - done_j) \max_{d' \in \mathcal{D}} Q_{\text{final}}\left(z'_j, d', c^{\mu}(z'_j; \theta^{-}); \omega^{-}\right),
    \label{eq:td_target_final}
\end{equation}
where \(z'_j\) denotes the next state of sample \(j\), and \(\theta^{-}\), \(\omega^{-}\) denote the target network parameters.
The expert parameters \(\omega\) are updated by minimizing the weighted Mean Squared Error (MSE) over the batch:
\begin{equation}
    \mathcal{L}_Q(\omega) = \frac{1}{B_{batch}} \sum_{j=1}^{B_{batch}} w_j \left( y_j - Q_{\text{final}}(z_j, d_j, c_j; \omega) \right)^2.
    \label{eq:loss_q_final}
\end{equation}
Crucially, to enforce functional specialization, gradients are backpropagated solely to the selected experts. For a specific expert \(i\) and sample \(j\), the gradient contribution is masked as:
\begin{equation}
    \nabla_{\omega_i} \mathcal{L}^{(j)}_Q = \mathbb{I}(i \in \Omega_j) \cdot \alpha_i(z_j) \cdot \frac{\partial \mathcal{L}^{(j)}_Q}{\partial Q_{\text{final}}} \cdot \nabla_{\omega_i} Q_i(z_j, d_j, c_j; \omega_i),
\end{equation}
where \(\mathbb{I}(\cdot)\) denotes the indicator function. This masking mechanism prevents parameter erosion in dormant experts (\(i \notin \Omega_j\)), ensuring they retain specialization for their specific motion manifolds.

\textbf{Continuous Action Network Update:}
The continuous action network \(\theta\) is updated via the Deterministic Policy Gradient (DPG) method. The loss function averages the Q-values over the batch:
\begin{equation}
    \mathcal{L}_{\mu}(\theta) = -\frac{1}{B_{batch}} \sum_{j=1}^{B_{batch}} Q_{\text{final}}(z_j, d_j, c^{\mu}(z_j; \theta); \omega).
    \label{eq:loss_mu}
\end{equation}
Here, the fixed critic network serves as a differentiable proxy, guiding \(\theta\) to output optimal modulation parameters.

Finally, the target networks are updated via soft replacement: \(\phi^{-} \leftarrow \tau \phi + (1-\tau) \phi^{-}\), where \(\phi \in \{\omega, \theta\}\).

\begin{algorithm}[t]
\caption{Learning Procedure for DHEA-MECD}
\label{alg:mecdm_update}
\begin{algorithmic}[1]
\State \textbf{Input:} PER Buffer $\mathcal{B}$, Batch Size $B_{batch}$
\State \textbf{Initialize:} Expert parameters $\omega$, Continuous policy $\theta$, Target networks $\omega^{-} \leftarrow \omega, \theta^{-} \leftarrow \theta$

\While{training}
    \State Sample a minibatch of $B_{batch}$ transitions from $\mathcal{B}$ by prioritized weights $w_j$
    \State Extract latent states $z_j$ and next states $z'_j$ via information extraction framework (described in Section \ref{Section:5-1})
    
    \For{each transition $j$ in minibatch}
        \State Compute target continuous action $c'_j $ by Eq. (\ref{eq:continuous_policy})
        \State Selected experts $\Omega'_j$ for $z'_j$ by Eq. (\ref{eq:gate_score})
        \State Compute TD target $y_j$ using Eq. (\ref{eq:td_target_final})
    \EndFor
    
    \State Compute experts network Loss $\mathcal{L}_Q(\omega)$ by Eq. (\ref{eq:loss_q_final})
    \State Update $\omega$ via gradient descent

    \State Compute continuous network Loss $\mathcal{L}_{\mu}(\theta)$ by Eq. (\ref{eq:loss_mu})
    \State Update $\theta$ via gradient descent 

    \State Soft update target networks: $\omega^{-} \leftarrow \tau \omega + (1-\tau) \omega^{-}$, $\theta^{-} \leftarrow \tau \theta + (1-\tau) \theta^{-}$
    \State Update information extraction framework parameter
    \State Update transition priorities in $\mathcal{B}$ based on TD error $\delta_j$  by Eq.~(\ref{eq:td_error})
\EndWhile
\end{algorithmic}
\end{algorithm}

\section{Proposed Tracking Algorithm Based on DHEA-MECD Algorithm}\label{Section:6}

Built upon the proposed DHEA-MECD algorithm, this section presents an AUV-based multi-target tracking algorithm designed for underwater environments with high-dimensional features.
As noted earlier, our AUV multi-target tracking framework adopts an EI architecture, which enables efficient target acquisition in complex marine environments through a hierarchical perception–decision–execution process. 
Within this pipeline, the information extraction framework described in Section~\ref{Section:5} is adopted to distill compact latent representations from high-dimensional environmental states, while the proposed multi-expert collaborative decision mechanism subsequently produces the hybrid (discrete–continuous) action required for robust AUV tracking.

\subsection{Reward Design}\label{Section:6-1}

To enable stable, efficient, and safe AUV multi-target tracking in underwater environments with high-dimensional features, we design a structured reward function $r$ as shown in Eq.~(\ref{rt}):

\begin{equation}
    r(t) = r_d(t) + r_g(t) + r_c(t) + r_b(t),
\label{rt}
\end{equation}
where $r_d(t)$ is the distance reward, $r_g(t)$ is the goal reward, $r_c(t)$ is the collision-avoidance reward, and $r_b(t)$ is the workspace adherence reward. 

\textbf{Distance Reward} $r_d$
encourages the AUV to continuously approach the target at every time step. The reward is defined as Eq.~(\ref{rd}):
\begin{equation}
    r_d(t) = \alpha_d \big( d_{a,j}(t-1) - d_{a,j}(t) \big),
\label{rd}
\end{equation}
where $d_{a,j}(t) = \|p_a(t) - p_j(t)\|_2$ denotes the Euclidean distance between the AUV and the target at time $t$ , $d_{a,j}(t-1)$ is the distance at  $t-1$ time , and $\alpha_d > 0$ is a scaling coefficient that regulates the magnitude of the distance reward.

\textbf{Goal Reward} $r_g$ 
provides a sparse but significant terminal signal when the AUV successfully reaches the target. This reward term is designed to reinforce the successful completion of the mission and steer the policy toward global convergence. The  reward is defined as Eq.~(\ref{rg}):
\begin{equation}
    r_g(t) = \alpha_g \cdot \mathbb{I}_{goal}(t),
\label{rg}
\end{equation}
where $\mathbb{I}_{goal}(t) \in \{0, 1\}$ is a binary indicator function that yields 1 if the target is successfully tracked at time $t$ and 0 otherwise, and $\alpha_g > 0$ denotes the scaling coefficient that scales the contribution of the incentive for terminal task achievement.

\textbf{Collision-Avoidance Reward} $r_c$ 
serves as a negative incentive to mitigate the occurrence of collisions between the AUV and environmental obstacles. The  reward is defined as Eq.~(\ref{rc}):
\begin{equation}
r_c(t) = -\alpha_c \cdot \mathbb{I}_{\text{collision}}(t),
\label{rc}
\end{equation}
where $\mathbb{I}_{\text{collision}}(t) \in \{0, 1\}$ is a binary indicator function that yields 1 if a collision is detected at time $t$ and 0 otherwise. The term $\alpha_c > 0$ denotes the scaling coefficient that modulates the impact of collision-related costs .

\textbf{Workspace Adherence Reward} $r_b$ 
is incorporated to enforce spatial constraints on the AUV's movements. This term assigns a boundary cost whenever the AUV exceeds the boundaries, ensuring the agent remains within the valid workspace. The reward is defined as Eq.~(\ref{rb}):
\begin{equation}
    r_b(t) = -\alpha_b \cdot\mathbb{I}_{\text{boundary}}(t) ,
\label{rb}
\end{equation}
where $\mathbb{I}_{\text{boundary}}(t) \in \{0, 1\}$ is a binary indicator function that yields 1 if the AUV exceeds the workspace boundaries and 0 otherwise. The coefficient $\alpha_b $ scales this negative incentive to effectively confine the AUV to the permissible domain.

\subsection{Multi-Target Tracking Algorithm Based on DHEA-MECD}\label{Section:5-2-2}
\hfill

The complete execution of the the proposed multi-target tracking algorithm based on the EI-driven DHEA-MECD summarized in Algorithm~\ref{alg:tracking_scheme}.
Specifically, Lines~1--6 initialize the networks and transform the raw state $s_t$ into a latent representation $z_t$ through semantic decomposition and multi-head encoder and multi-head self-attention-based fusion.
Subsequently, Lines~7--14 generate the hybrid action by dynamically selecting a Top-$k$ expert subset $\Omega_t$ and fusing their outputs to determine both continuous action $c_t$ and discrete action $d_t$.
Finally, Lines~15--19 execute the assembled action $a_t$, compute the multi-target tracking reward $r_t$, and perform joint optimization of the perception and decision modules via Algorithm~\ref{alg:mecdm_update} to ensure robust tracking performance.

\begin{algorithm}[t]
\caption{Proposed Tracking Algorithm based on DHEA-MECD}
\label{alg:tracking_scheme}
\begin{algorithmic}[1]
\State \textbf{Input:} Environmental state $s_t$, target expert count $k$, scaling coefficients $\alpha$
\State \textbf{Initialize:} Replay Buffer $\mathcal{B}$, hierarchical encoder parameters, self-attention weights, continuous and expert networks

\While{task is active}
    \State Observe raw observation $s_t$ and normalize to $[0, 1]$
    \State Decompose $s_t$ into four semantic subspaces: $\{s_{\text{sp}}, s_{\text{mot}}, s_{\text{bar}}, s_{\text{noise}}\}$;
    \State Distill modality-specific features using specialized multi-head encoders
    \State Model cross-modal dependencies via multi-head self-attention to generate latent state $z_t$
    
    \State Generate continuous action  $c_t$ by Eq.~(\ref{eq:continuous_policy})
    \State Calculate matching scores and select top-$k$ expert subset $\Omega_t$ Eq.~(\ref{eq:gate_score})
    \For{each selected expert $i \in \Omega_t$}
        \State Calculate expert weight $\alpha_i$ by Eq.~(\ref{eq:sparse_softmax})
        \State Compute  expert action-value $Q_i(z_t, d, c_t)$ 
    \EndFor
    \State Synthesize global action-value $Q_{\text{final}}$ by Eq.~(\ref{eq:qfinal})
    \State Select discrete action $d_t$ by Eq.~(\ref{finaldt})
    \State Assemble hybrid action $a_t = (d_t, c_t)$
    \State Execute $a_t$ via AUV propulsion system 
    \State Obtain reward $r_t$ by Eq.~(\ref{rt}) and get next state $s_{t+1}$
    \State Append transition $e_t = (z_t, d_t, c_t, r_t, z_{t+1}, done_t)$ to prioritized buffer $\mathcal{B}$
    \State Perform joint optimization of encoders and policies via Algorithm \ref{alg:mecdm_update}
\EndWhile
\end{algorithmic}
\end{algorithm}

\section{Evaluations}\label{Section:7}
This section presents comprehensive comparative evaluations between the proposed approach and several mainstream DRL-driven  approaches. 

\subsection{Simulation setup}\label{Section:7-1}
All the simulations are conducted on a MacBook Air (2024) equipped with an Apple M3 processor and 16 GB memory. In the simulation, both the AUV and the target entities are abstracted as dynamic particles evolving within a three-dimensional continuous space. 
All the parameters in the simulations are detailed in Table. \ref{table:simulation_parameters}.

\begin{table}[H]
\centering
\caption{Simulation parameters}
\label{table:simulation_parameters}
\begin{tabular}{ccc}
\hline
\textbf{Parameter} & \textbf{Description} & \textbf{Value} \\ 
\hline
$T_{\max}$ & Maximum steps per episode & 800 \\
$N_{\text{episodes}}$ & Total training episodes & 10,000 \\

$N_{\text{expert}}$ & Number of experts & 5 \\
$k$ & Selected experts per step & 2 \\

$\alpha$ & Expert network learning rate & 0.001 \\
$\beta$ & Continuous network learning rate & 0.001 \\
$\gamma$ & Discount factor & 0.95 \\
$T_{\text{update}}$ & Target network update frequency & 200 \\
$B_{batch}$ & Batch size & 64 \\
$M$ & Replay buffer capacity & 200,000 \\

$\epsilon_0$ & Initial exploration rate & 1.0 \\
$\lambda_\epsilon$ & Exploration decay rate & 0.0002 \\
$\|\nabla\|_{\max}$ & Gradient clipping norm & 1.0 \\
\hline
\end{tabular}
\end{table}

\subsection{Results}\label{Section:6-2}
To evaluate the effectiveness of the proposed DHEA-MECD in multi-target tracking, we conduct a comprehensive comparison against a suite of mainstream DRL algorithm-driven approaches, including PPO~\cite{ppo}, PDQN~\cite{pdqn}, SAC~\cite{sac}, DDPG~\cite{ddpg}, TRPO~\cite{trpo}, AC~\cite{ac}, and DQN~\cite{dqn}, in terms of convergence speed,  trajectory length, tracking accuracy, collision rate, respectively.

\begin{figure*}
\centering

\subfloat[With 9-dimensional features]{%
  \includegraphics[width=0.49\textwidth]{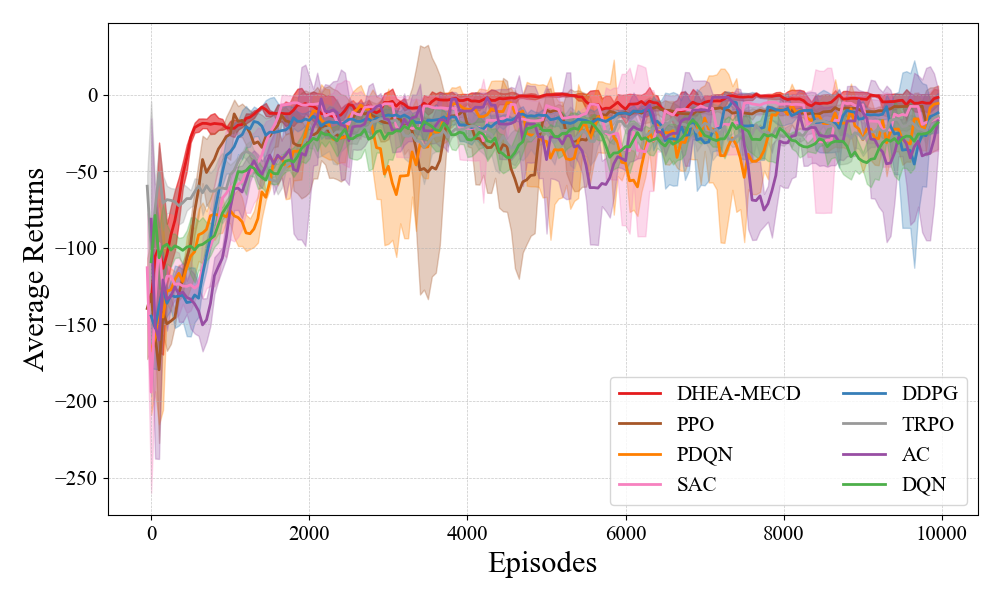}%
  \label{fig:4a}%
}\hfill
\subfloat[With 18-dimensional features]{%
  \includegraphics[width=0.49\textwidth]{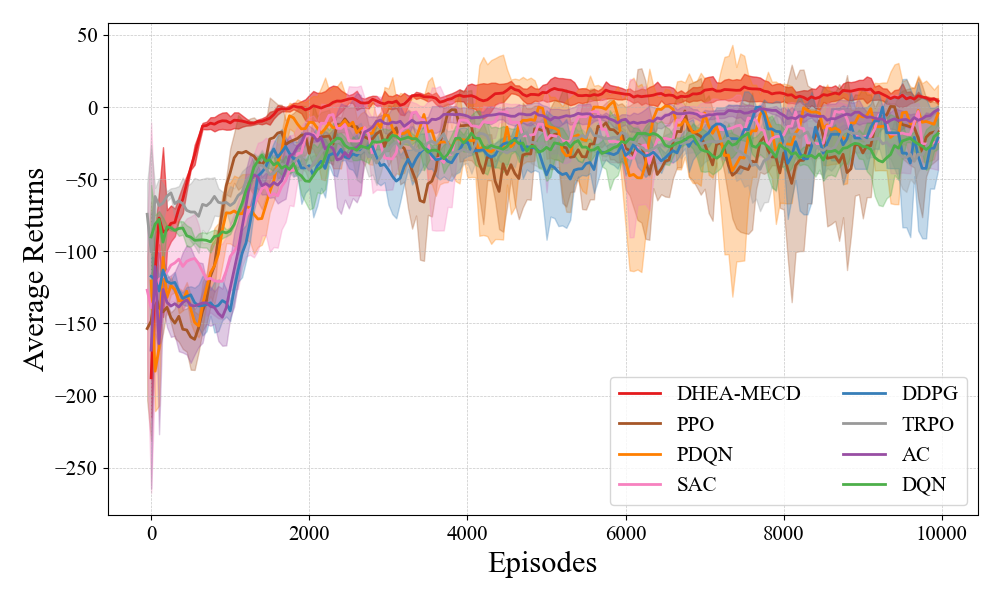}%
  \label{fig:4b}%
}

\subfloat[With 27-dimensional features]{%
  \includegraphics[width=0.49\textwidth]{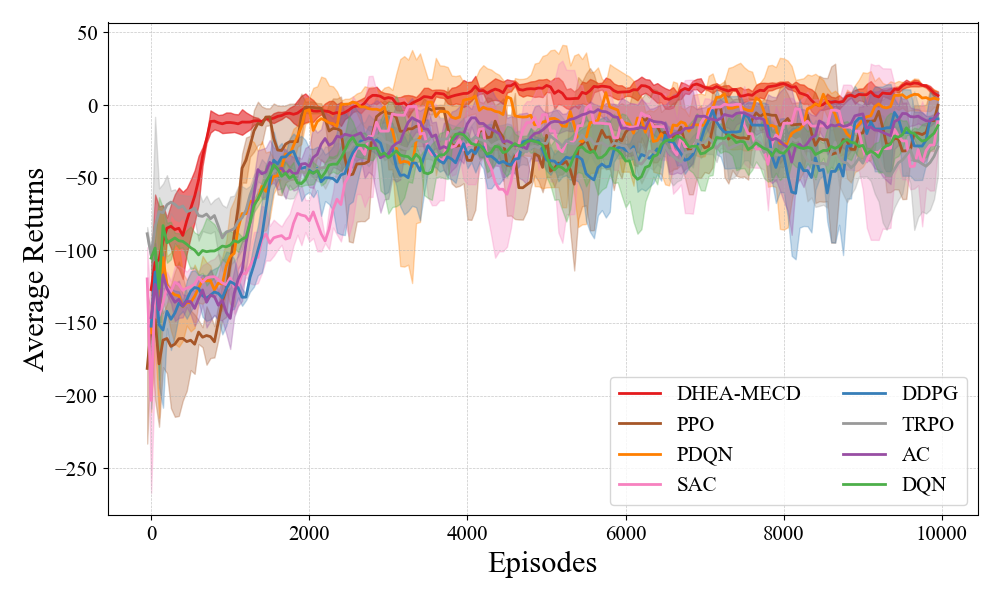}%
  \label{fig:4c}%
}\hfill
\subfloat[With 36-dimensional features]{%
  \includegraphics[width=0.49\textwidth]{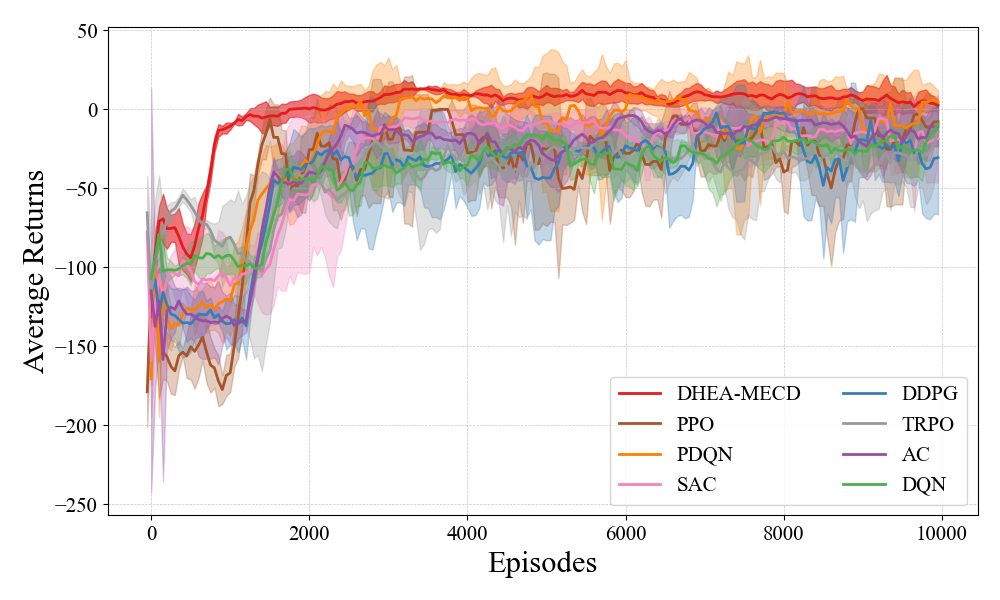}%
  \label{fig:4d}%
}

\caption{Convergence speed comparison}
\label{fig:4}

\end{figure*}

\makeatletter
\def\blfootnote{\gdef\@thefnmark{}\@footnotetext}
\makeatother

\blfootnote{\textsuperscript{$\dagger$} Detailed specifications of the dimensional features are provided in the Appendix, as part of the ``Appendices'' section.}

\textbf{1) Convergence Speed:
}
For evaluating the convergence speed, we conduct multiple experimental trials under varying state-feature dimensional environments\textsuperscript{$\dagger$}  and evaluates the convergence of different DRL algorithms using averaged returns with a 96\% confidence interval. 

As the results in Fig.~\ref{fig:4}, the proposed DHEA-MECD consistently achieves faster convergence than other algorithms in most scenarios. Notably, as the dimensionality of the state features increases, the performance gap between DHEA-MECD and the compared algorithms becomes increasingly pronounced.
In particular, the compared PPO, SAC, DDPG, DQN) exhibit significant performance variability in high-dimensional spaces, primarily due to their limited capability in structured feature extraction. Consequently, they struggle with redundant and noisy representations, leading to unstable optimization.

In contrast, DHEA-MECD demonstrates markedly faster early-stage convergence. This efficiency stems from the multi-head encoder and attention architecture, which decomposes high-dimensional observations into coherent latent manifolds. By distilling compact embeddings $z_t$, this structural bias effectively filters noise and redundancy, facilitating more efficient exploration than compared approaches.
Furthermore, the Top-$k$ expert selection strategy in DHEA-MECD accelerates policy search by restricting gradient propagation to relevant sub-networks.
Unlike the monolithic architectures of the compared DRL algorithms, our proposed DHEA-MECD minimizes parameter interference in complex hybrid action spaces.
In later training stages, DHEA-MECD maintains a smooth trajectory with minimal oscillation, supported by PER mechanism. By focusing gradient updates on high-impact transitions, PER enhances value estimation stability. 
Collectively, the synergy of structured representation, sparse collaboration, and prioritized replay ensures robust convergence, enabling reliable tracking performance in dynamic, high-dimensional underwater environments.

\begin{table*}[t]
\centering
\small
\caption{Trajectory length distribution comparison.}
\label{tab:length comparison}
\resizebox{\linewidth}{!}{
\begin{tabular}{c||cccc|cccc|cccc|cccc}
\toprule
\textbf{Dimensions} & \multicolumn{4}{c|}{\textbf{With 9-dimensional features}} & \multicolumn{4}{c|}{\textbf{With 18-dimensional features}} & \multicolumn{4}{c|}{\textbf{With 27-dimensional features}} & \multicolumn{4}{c}{\textbf{With 36-dimensional features}} \\
\cmidrule(lr){2-5} \cmidrule(lr){6-9} \cmidrule(lr){10-13} \cmidrule(lr){14-17}
\textbf{Length(m)}
& 200-300 & 300-350 & 350-400 & 400-500 & 200-300 & 300-350 & 350-400 &  400-500 & 200-300 & 300-350 & 350-400 & 400-500 & 200-300 & 300-350 & 350-400 & 400-500 \\
\midrule
PPO & 35.2\% & 24.2\% & 27.1\% & 13.5\% & 38.1\% & 24.2\% & 25.1\% & 12.6\% & 42.3\% & 26.9\% & 19.9\% & 10.9\% & 44.6\% & 26.7\% & 18.8\% & 9.9\% \\
PDQN  & 44.4\% & 24.8\% & 19.8\% & 10.9\% & 46.4\% & 26.0\% & 17.7\% & 9.8\% & 49.7\% & 27.9\% & 14.9\% & 7.5\% & 53.8\% & 25.4\% & 14.2\% & 6.6\% \\
SAC      & 11.7\% & 13.2\% & 9.2\% & 65.9\% & 12.9\% & 13.6\% & 9.0\% & 64.5\% & 13.2\% & 15.3\% & 8.7\% & 62.7\% & 15.7\% & 16.1\% & 8.4\% & 59.9\% \\
DDPG      & 8.2\% & 10.9\% & 17.3\% & 63.6\% & 8.6\% & 11.9\% & 17.0\% & 62.5\% & 9.0\% & 12.3\% & 17.0\% & 61.7\% & 10.3\% & 13.5\% & 16.1\% & 60.1\% \\
TRPO      & 4.9\% & 7.3\% & 17.1\% & 70.7\% & 5.3\% & 7.7\% & 16.9\% & 70.0\% & 5.7\% & 8.2\% & 16.4\% & 69.7\% & 6.3\% & 9.0\% & 15.9\% & 68.8\% \\
AC    & 4.8\% & 13.0\% & 18.8\% & 63.3\% & 5.2\% & 13.5\% & 18.6\% & 62.7\% & 5.6\% & 14.1\% & 18.4\% & 61.9\% & 6.1\% & 14.6\% & 18.6\% & 60.7\% \\
DQN    & 5.8\% & 7.2\% & 14.5\% & 72.5\% & 6.6\% & 8.1\% & 14.2\% & 71.1\% & 7.5\% & 9.0\% & 13.9\% & 69.7\% & 8.5\% & 9.7\% & 13.6\% & 68.2\% \\
\midrule
\textbf{DHEA-MECD} & \textbf{56.5\%} & \textbf{21.7\%} & \textbf{16.3\%} & \textbf{5.4\%} & \textbf{59.1\%} & \textbf{22.7\%} & \textbf{13.6\%} & \textbf{4.5\%} & \textbf{63.1\%} & \textbf{22.5\%} & \textbf{10.8\%} & \textbf{3.6\%} & \textbf{66.7\%} & \textbf{24.1\%} & \textbf{6.5\%} & \textbf{2.8\%} \\
\bottomrule

\end{tabular}
}
\end{table*}

\begin{figure}[t!]
    \centering
               \includegraphics[width=1\linewidth]{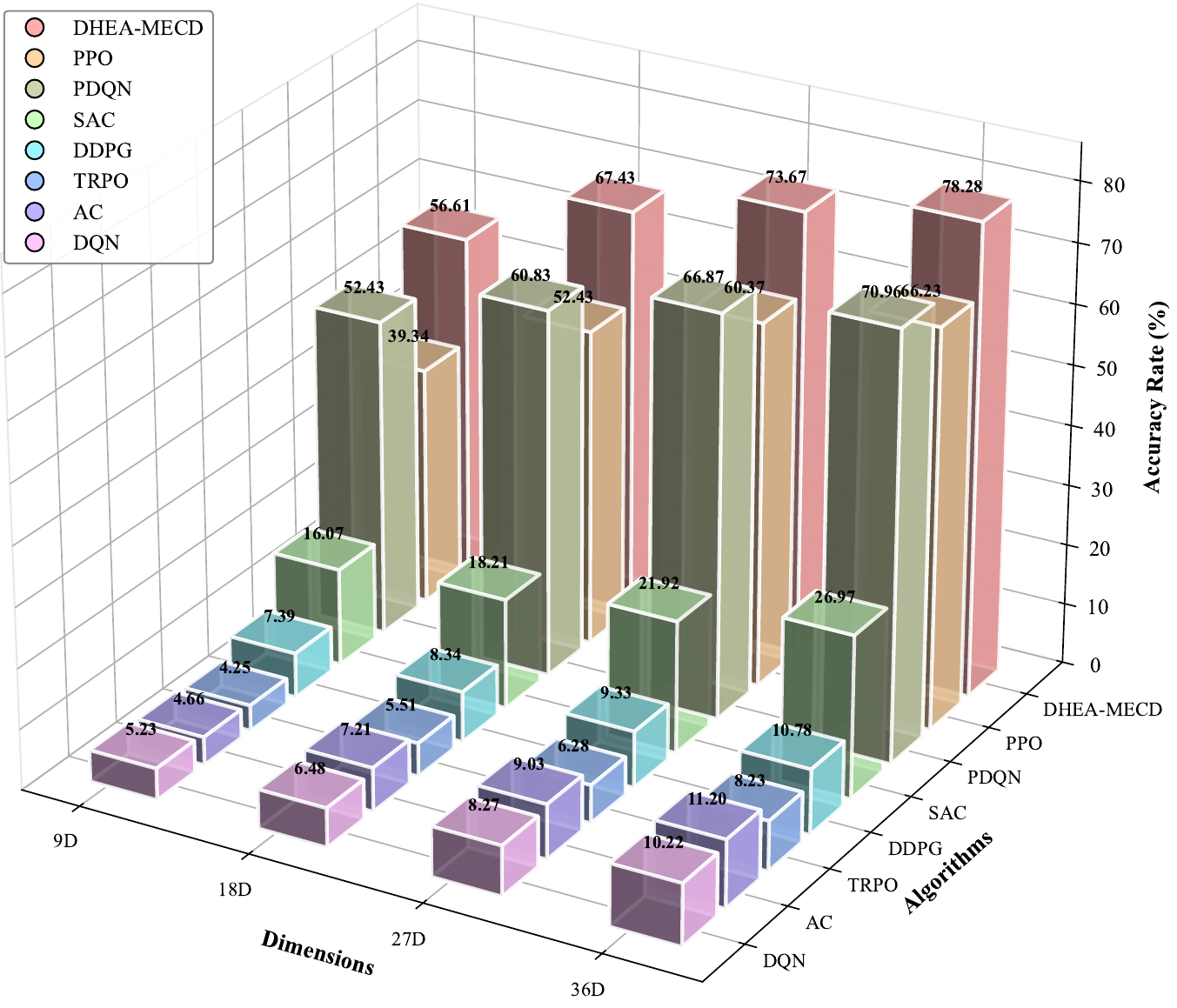}
    \caption{Accuracy rate comparison }
    \label{fig6}
\end{figure}

\begin{figure}[t!]
    \centering
               \includegraphics[width=1\linewidth]{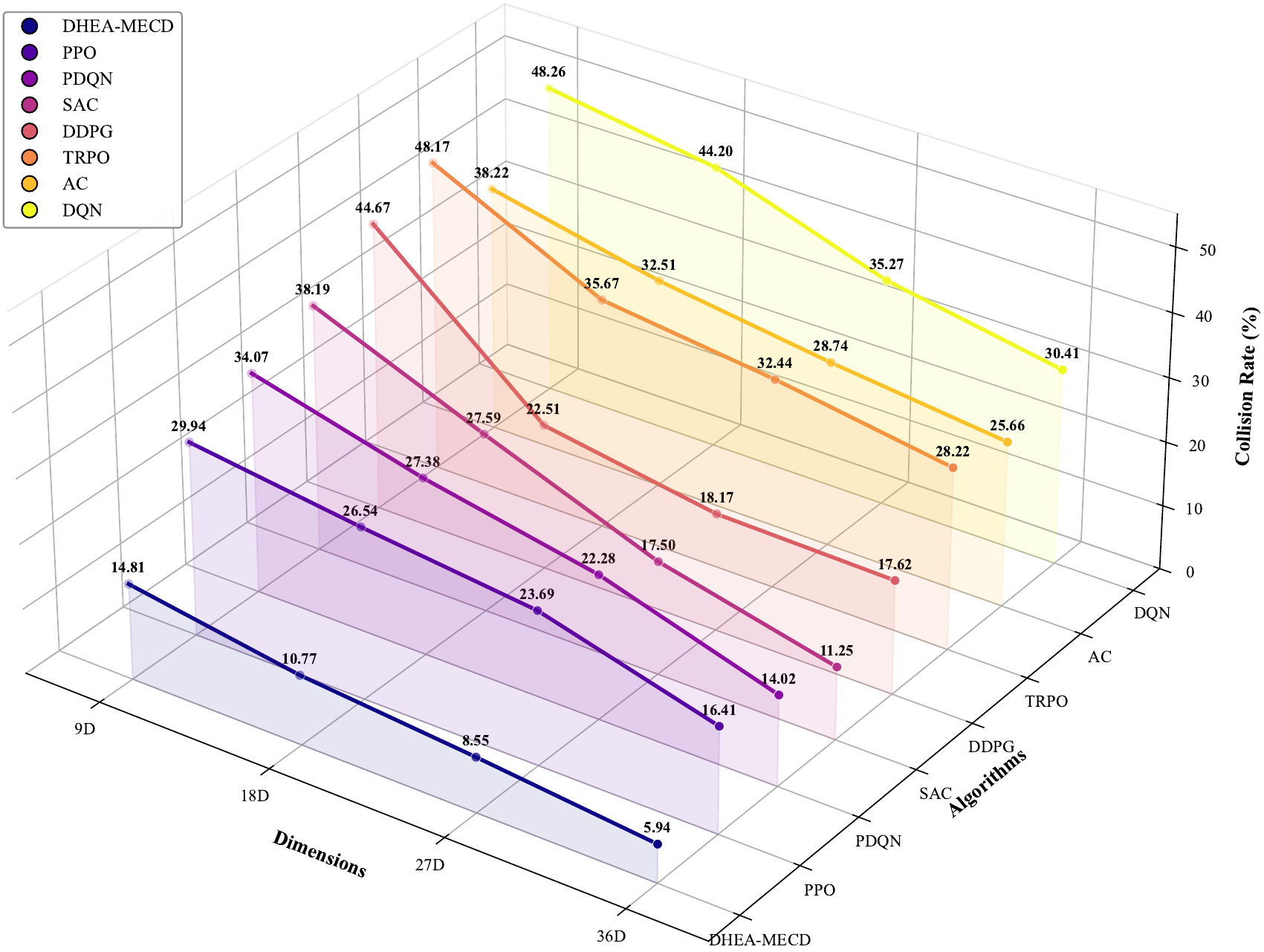}
    \caption{Collision rate comparison}
    \label{fig7}
\end{figure}

\begin{figure}[t!]
    \centering
    \subfloat[Ablation with 18-dimensional features]{%
        \includegraphics[width=1.0\linewidth]{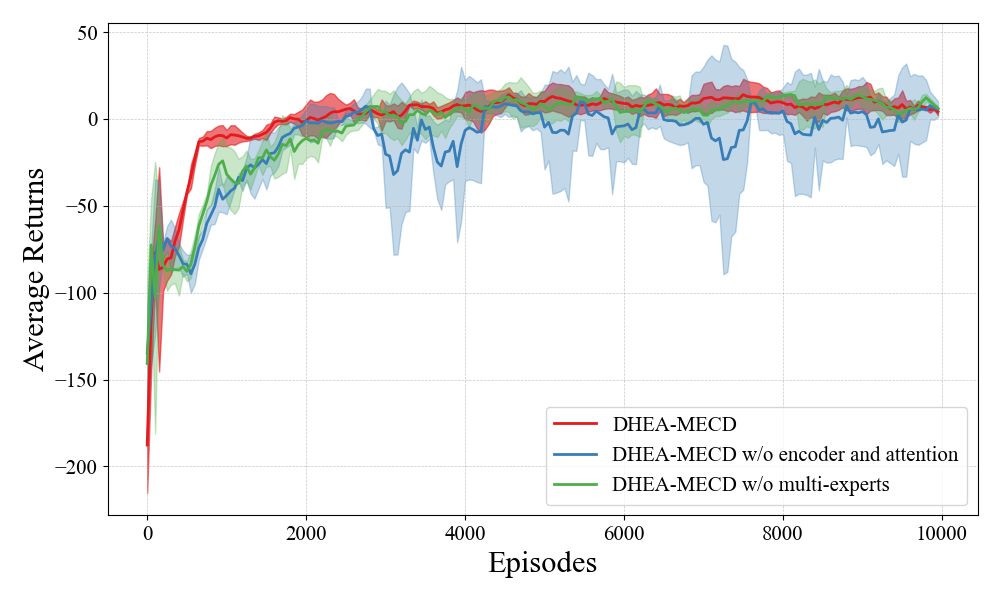}%
        \label{fig:ablation18}
    } \\ 
    
    \vspace{1em} 
    
    \subfloat[Ablation with 36-dimensional features]{%
        \includegraphics[width=1.0\linewidth]{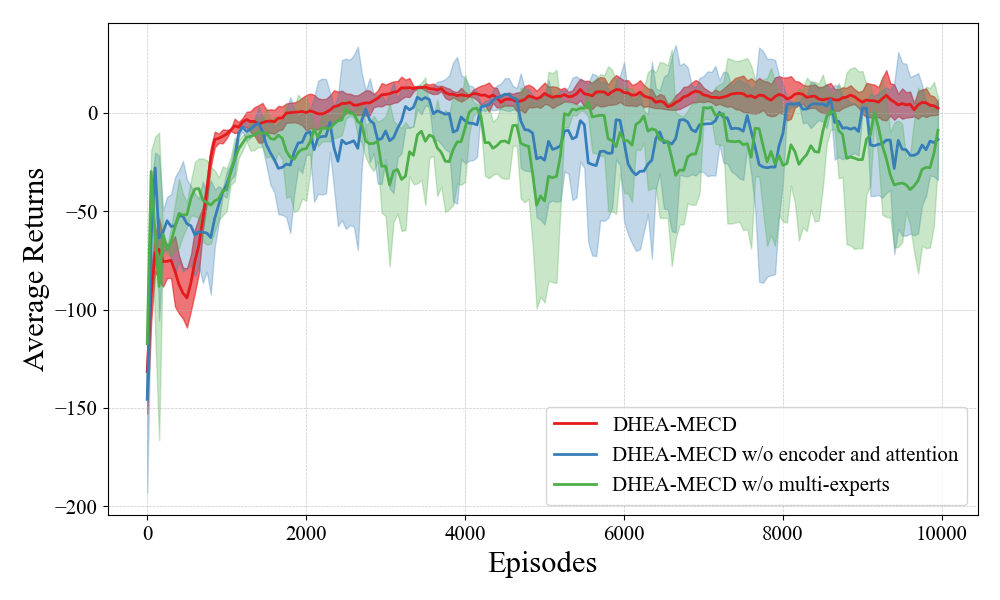}%
        \label{fig:ablation36}%
    }

    \caption{Ablation study} 
    \label{fig:ablation_comparison}
\end{figure}

\begin{figure*}[t]
    \centering
    \subfloat[Top view of tracking process]{%
        \begin{minipage}{0.32\linewidth}
            \centering
            \includegraphics[width=\linewidth]{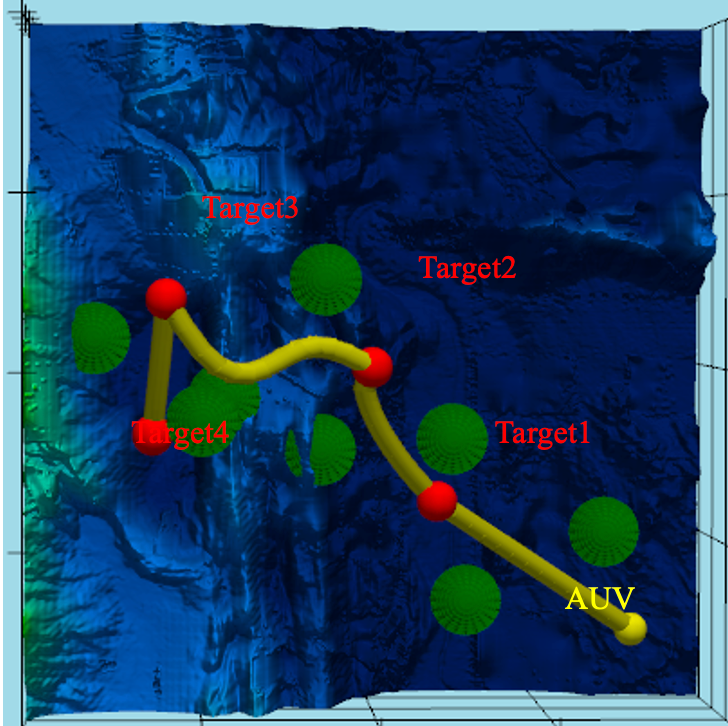}
            \label{fig:sim1}
        \end{minipage}
    }\hfill 
    \subfloat[Side view 1 of tracking process]{%
        \begin{minipage}{0.32\linewidth}
            \centering
            \includegraphics[width=\linewidth]{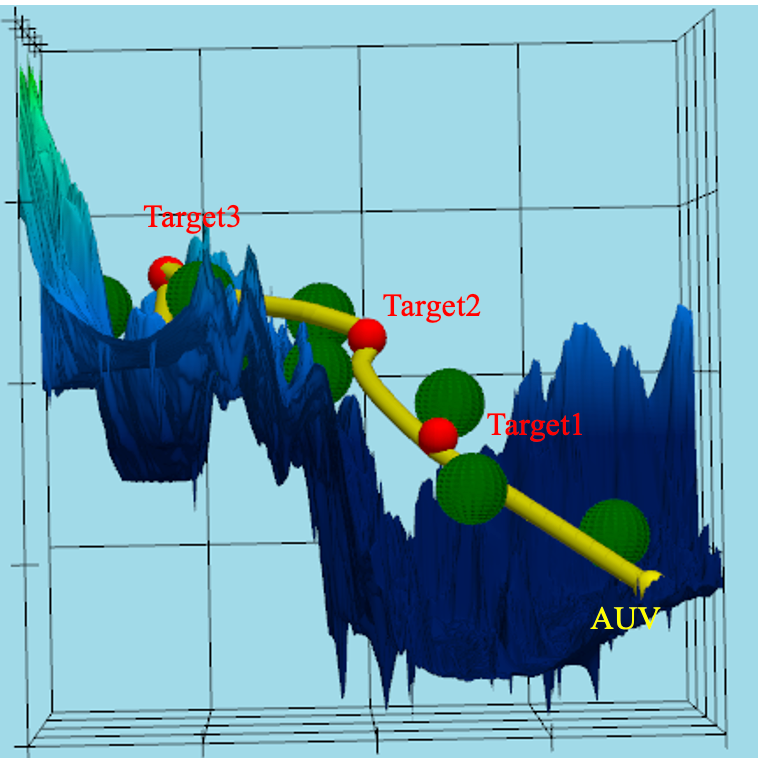}
            \label{fig:sim2}
        \end{minipage}
    }\hfill
    \subfloat[Side view 2 of tracking process]{%
        \begin{minipage}{0.32\linewidth}
            \centering
            \includegraphics[width=\linewidth]{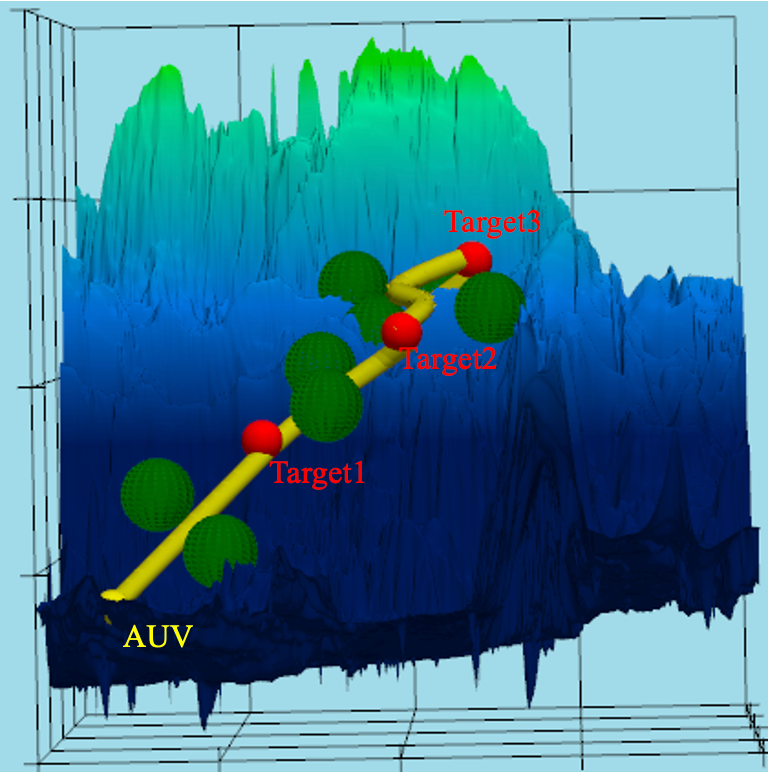}
            \label{fig:sim3}
        \end{minipage}
    }
    \caption{Visualization results.}
    \label{fig:10}
\end{figure*}

\textbf{2) Trajectory Length:}
The trajectory length, defined as the cumulative distance traveled by the AUV during the target-following process, is adopted as a key metric for evaluating tracking efficiency.
Table~\ref{tab:length comparison} showcases the distribution of the tracking path required by different algorithms to successfully track the target under varying feature dimensions. 
For all the scenarios (with different feature dimensions), DHEA-MECD consistently achieves the highest concentration of successful episodes in the 200–300m interval.
Crucially, this performance advantage becomes increasingly pronounced as dimensions rise from 9 to 36. While the
other compared approaches efficiency degrades significantly in high-dimensional environments, DHEA-MECD maintains a high-quality performance. This results from its structured representation learning and multi-expert collaborative decision mechanism, which effectively mitigate the complex dimensionality and minimize redundant motion.
The superior trajectory achieved by DHEA-MECD primarily arises from the structural advantages of its hierarchical architecture. The double-head encoder--attention module distills high-dimensional observations into a noise-suppressed and semantically aligned latent representation, enabling more accurate inference of spatial--temporal dependencies and reducing superfluous corrective actions.
Meanwhile, the motion-stage-aware Top-$k$ expert selection mechanism concentrates decision-making on the most stage-relevant experts, while the hybrid (discrete--continuous) action execution produces smoother and more directionally consistent trajectories.
Collectively, these mechanisms allow DHEA-MECD to transform richer environmental information into decisively goal-oriented behaviors, thereby yielding shorter and more direct tracking paths across all feature-dimension environments.

\textbf{3) Tracking Accuracy:
}Fig.~\ref{fig6} illustrates the success rate comparison of different algorithms under varying observation feature dimensions. 
As the feature dimension increases from 9 to 36, the proposed DHEA-MECD consistently achieves the highest success rate among all compared methods. Notably, its performance advantage becomes increasingly pronounced in high-dimensional environments, demonstrating strong scalability and robustness.

In contrast, compared algorithms such as PPO, SAC, and DDPG exhibit a clear degradation in success rate as the observation dimension increases. 
This decline performance indicates their limited capability to handle complex and high-dimensional state spaces, where redundant features and noisy observations significantly impair policy learning. 
Particularly in the 36-dimensional environment, the compared methods suffer from substantial performance drops, highlighting their vulnerability to the curse of dimensionality.

Diverging from conventional single-policy methods, DHEA-MECD explicitly decouples structured state representation learning from hybrid action decision-making through a multi-expert collaboration mechanism. 
This design enables it to maintain reliable target-tracking performance even in challenging high-dimensional environments.

\textbf{4) Collision Rate:}
Collision avoidance is a critical requirement in multi-target tracking tasks, particularly in underwater environments with high-dimensional features.

Fig.~\ref{fig7} presents the collision rate comparison of different algorithms under varying observation feature dimensions. In low-dimensional environments, all algorithms exhibit relatively low collision rates. However, as the feature dimension increases, the collision rates of compared methods rise noticeably, indicating increasing difficulty in collision avoidance under complex and high-dimensional observations.

In contrast, the proposed DHEA-MECD consistently maintains the lowest collision rate across all environments with different feature dimensions. 
Notably, its collision rate decreases steadily from 9 to 36, demonstrating strong robustness and reliable obstacle avoidance performance even in high-dimensional environments. This highlights DHEA-MECD’s superior ability to learn stable and safe navigation behaviors under increased perception complexity.

In particular, the poor performance of  compared algorithms such as PPO, SAC, and DDPG can be attributed to their limited capability to coordinate motion planning and continuous control under high-dimensional observations. Within the same training duration, these methods tend to execute excessive turning or aggressive motion commands before learning effective obstacle avoidance strategies, thereby increasing collision risks. In contrast, DHEA-MECD explicitly decouples discrete motion strategy selection from continuous control parameter optimization, enabling more precise regulation of speed, heading, and braking behaviors when approaching obstacles.

\textbf{5) Ablation Study:}
To evaluate the individual contributions of the core components in the proposed approaches, ablation studies are conducted by comparing the full DHEA-MECD with two reduced variants: (1) DHEA-MECD without the multi-head encoder and multi-head self-attention module, and (2) DHEA-MECD without the multi-expert collaborative decision mechanism. The learning curves under different observation dimensions are displayed in Fig.~\ref{fig:ablation_comparison}.

As shown in Fig.~\ref{fig:ablation18}, at 18-dimensional observations, removing the multi-head encoder and attention module leads to noticeably slower convergence and larger performance fluctuations during the early training stage. This indicates that structured representation learning plays a critical role in accelerating early-stage policy optimization by extracting compact and informative state embeddings, thereby facilitating more efficient exploration.

In contrast, the variant without the multi-expert collaborative decision mechanism exhibits relatively comparable early learning behavior but suffers from increased oscillations and degraded performance in the mid-to-late training stages. This degradation suggests that expert collaboration is essential for stabilizing decision-making and reducing policy variance, particularly in complex hybrid action spaces.
Similar trends are observed in the more challenging 36-dimensional environment, as shown in Fig.~\ref{fig:ablation36}, where the absence of either component results in more pronounced performance degradation. Notably, the full DHEA-MECD consistently achieves higher average returns with smoother convergence trajectories across all training stages.

Finally, a grid-based scheme is employed to construct an underwater environment to emulate realistic operating conditions.
The DHEA-MECD-guided multi-target tracking process is visualized using the \texttt{pyvista} library.
As shown in Fig.~\ref{fig:10}, the red spheres denote the targets and the yellow spheres represent the AUV.
The solid curves illustrate the corresponding target and AUV trajectories under different tracking scenarios, as depicted in Figs.~\ref{fig:sim1}, Figs.~\ref{fig:sim2}, and Figs.~\ref{fig:sim3}.
The results indicate that the proposed DHEA-MECD algorithm enables the AUV to generate smooth, stable, and adaptive tracking trajectories in complex underwater environments, demonstrating its effectiveness and robustness under dynamic conditions.

\section{Conclusion}\label{Section:8}

In this paper, we introduce a hierarchical EI architecture to tackle the inherent complexities of AUV-based multi-target tracking in underwater environments with high-dimensional features. 
Based on our EI architecture, we propose the DHEA-MECD algorithm, which resolves the ``perception fragmentation" of complex marine environments by leveraging a multi-head encoder and multi-head self-attention module. Through a multi-expert collaborative decision mechanism with the motion-stage-aware Top-$k$ expert selection strategy, the proposed method effectively mitigates parameter interference and ensures high computational.
Extensive experimental evaluations demonstrate that DHEA-MECD consistently outperforms mainstream DRL algorithms (including PPO, PDQN, and DDPG) across varying feature dimensions. 
Specifically, our method exhibits superior sample efficiency, asymptotic stability, and tracking precision, with the performance gap widening notably as the observation dimensionality increases.
The ablation studies further confirm the synergistic necessity of structured representation learning and collaborative decision-making for robust underwater multi-target tracking.

In future work, we plan to extend this framework to multi-AUV cooperative tracking scenarios, investigating distributed expert coordination strategies to handle large-scale target clusters. 
Additionally, we aim to bridge the ``sim-to-real" gap by validating the proposed algorithm on physical AUV platforms under real-world hydrological conditions.


\appendices


\ifCLASSOPTIONcaptionsoff
  \newpage
\fi

\bibliographystyle{IEEEtran}
\bibliography{ref}

\vspace{-8ex}
\begin{IEEEbiography}[{\includegraphics[width=1in,height=1.25in,clip,keepaspectratio]{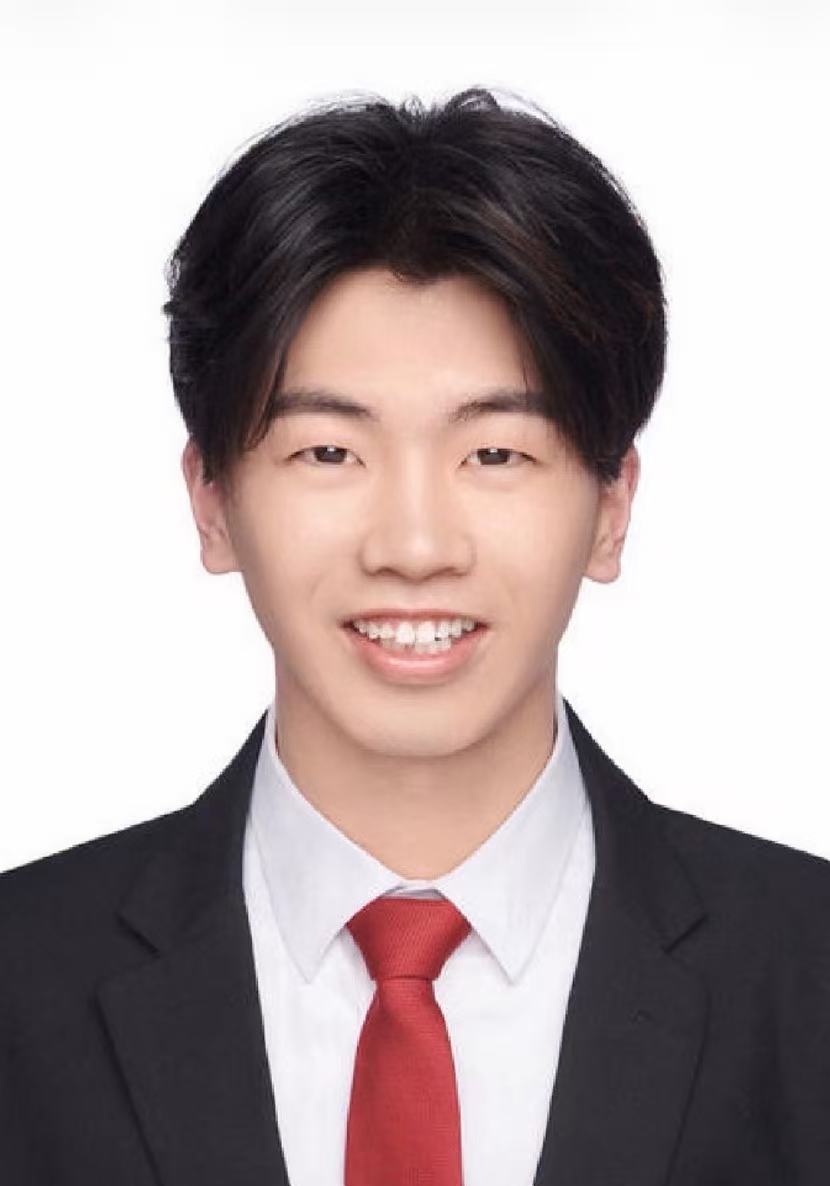}}]{Kai Tian}
	is currently working toward the bachelor’s degree with the Software College, Northeastern University, Shenyang, China. His research interests include embodied intelligence, deep reinforcement learning and machine learning.
\end{IEEEbiography}
\vspace{-3ex}

\begin{IEEEbiography}[{\includegraphics[width=1in,height=1.25in,clip,keepaspectratio]{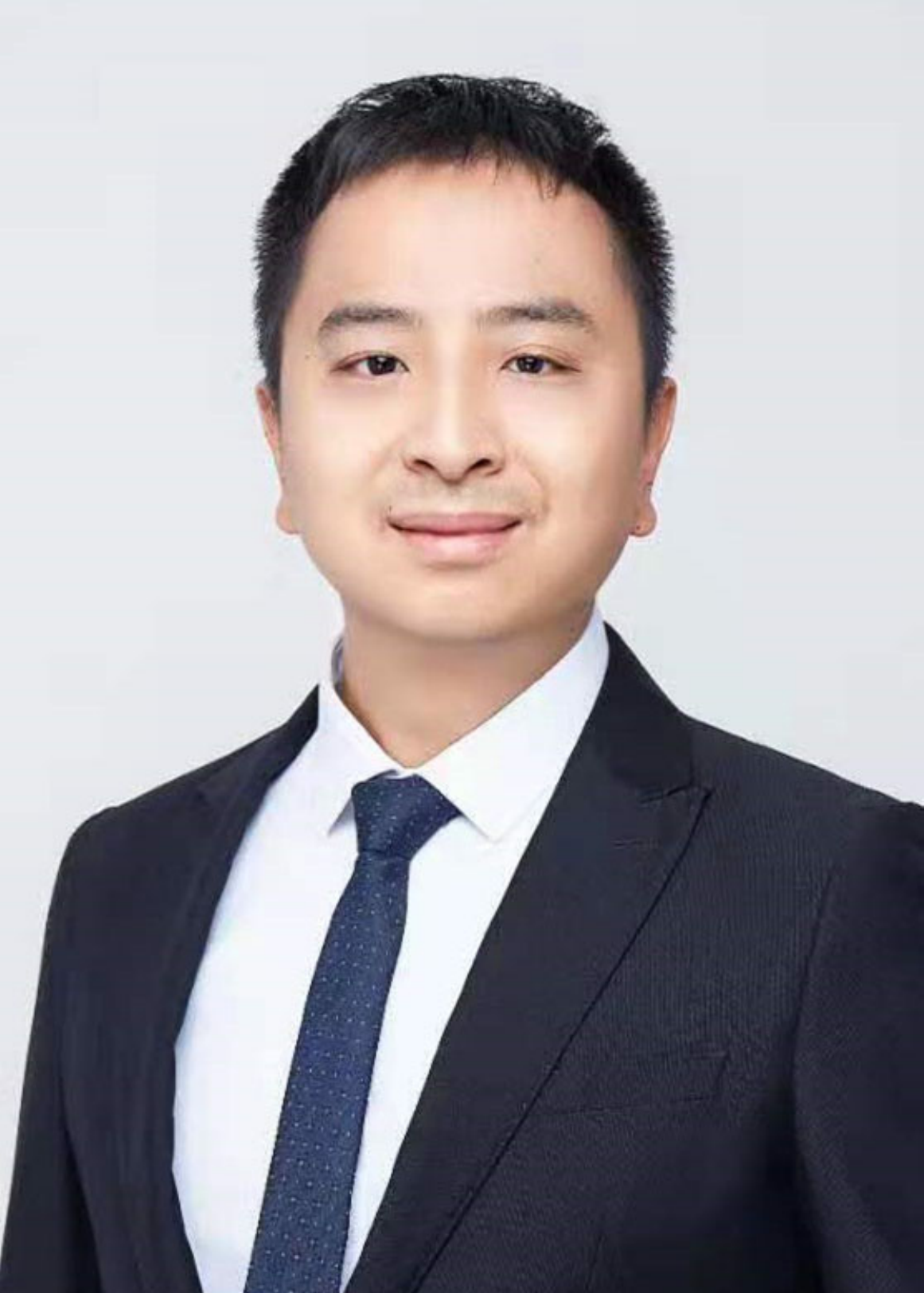}}]{Chuan Lin}
	[S'17, M'20] is currently an associate professor with the Software College, Northeastern University, Shenyang, China.
	He received the B.S. degree in Computer Science and Technology from Liaoning University, Shenyang, China in 2011, the M.S. degree in Computer Science and Technology from Northeastern University, Shenyang, China in 2013, and the Ph.D. degree in computer architecture in 2018.
	From Nove. 2018 to  Nove. 2020, he is a Postdoctoral Researcher with the School of Software, Dalian University of Technology, Dalian, China.
	His research interests include UWSNs, industrial IoT, software-defined networking.
\end{IEEEbiography}
\vspace{-3ex}
\begin{IEEEbiography}[{\includegraphics[width=1in,height=1.25in,clip,keepaspectratio]{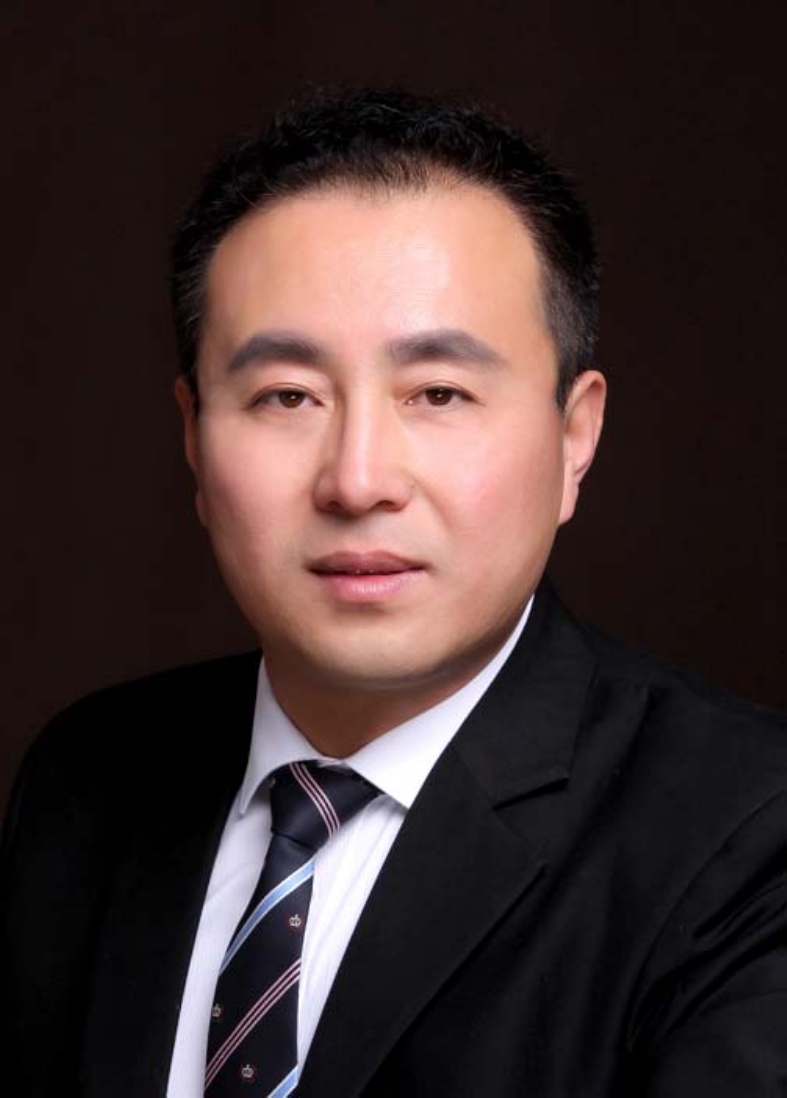}}]{Guangjie Han} [S’03-M’05-SM’18-F’22] is currently a Professor with the Department of Internet of Things Engineering, Hohai University, Changzhou, China. He received his Ph.D. degree from Northeastern University, Shenyang, China, in 2004. In February 2008, he finished his work as a Postdoctoral Researcher with the Department of Computer Science, Chonnam National University, Gwangju, Korea. From October 2010 to October 2011, he was a Visiting Research Scholar with Osaka University, Suita, Japan. From January 2017 to February 2017, he was a Visiting Professor with City University of Hong Kong, China. From July 2017 to July 2020, he was a Distinguished Professor with Dalian University of Technology, China. His current research interests include Internet of Things, Industrial Internet, Machine Learning and Artificial Intelligence, Mobile Computing, Security and Privacy. Dr. Han has over 500 peer-reviewed journal and conference papers, in addition to 160 granted and pending patents. Currently, his H-index is 65 and i10-index is 282 in Google Citation (Google Scholar). The total citation count of his papers raises above 15500+ times. Dr. Han is a Fellow of the UK Institution of Engineering and Technology (FIET). He has served on the Editorial Boards of up to 10 international journals, including the IEEE TII, IEEE TCCN, IEEE TVT, IEEE Systems, etc. He has guest-edited several special issues in IEEE Journals and Magazines, including the IEEE JSAC, IEEE Communications, IEEE Wireless Communications, Computer Networks, etc. Dr. Han has also served as chair of organizing and technical committees in many international conferences. He has been awarded 2020 IEEE Systems Journal Annual Best Paper Award and the 2017-2019 IEEE ACCESS Outstanding Associate Editor Award. He is a Fellow of IEEE.
\end{IEEEbiography}
\vspace{-8ex}

\begin{IEEEbiography}
[{\includegraphics[width=1in,height=1.25in,clip,keepaspectratio]{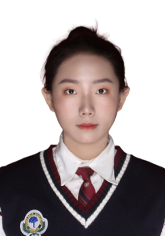}}]{Chen An}
 is currently pursuing the B.S. degree at the School of Computer Science and Engineering, Northeastern University, China. Her research interests include affective computing, emotional-expression-related hallucinations in large language models, and model interpretability.
\end{IEEEbiography}
\vspace{-8ex}
\begin{IEEEbiography}[{\includegraphics[width=1in,height=1.25in,clip,keepaspectratio]{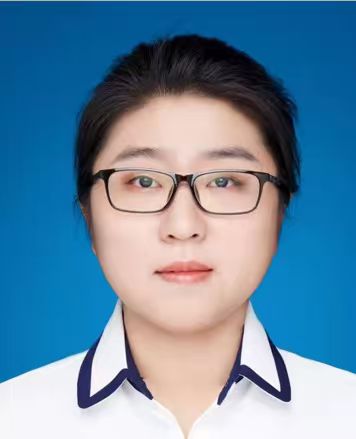}}]{Qian Zhu}
is currently an associate professor with the Software College, Northeastern University, Shenyang, China. She received the B.S. degree in Information and Computing Science from Northeastern University, Shenyang, China in 2006,  the M.S. degree in Operation Science and Control Theory from Northeastern University, Shenyang, China in 2008, and the Ph.D. degree in Communication and Information System from Northeastern University, China in 2018. Her research interests include artificial intelligence optimization algorithms, Unmanned Aerial Vehicle (UAV) technology and software development for applications.
\end{IEEEbiography}

\vspace{-8ex}

\begin{IEEEbiography}
[{\includegraphics[width=1in,height=1.25in,clip,keepaspectratio]{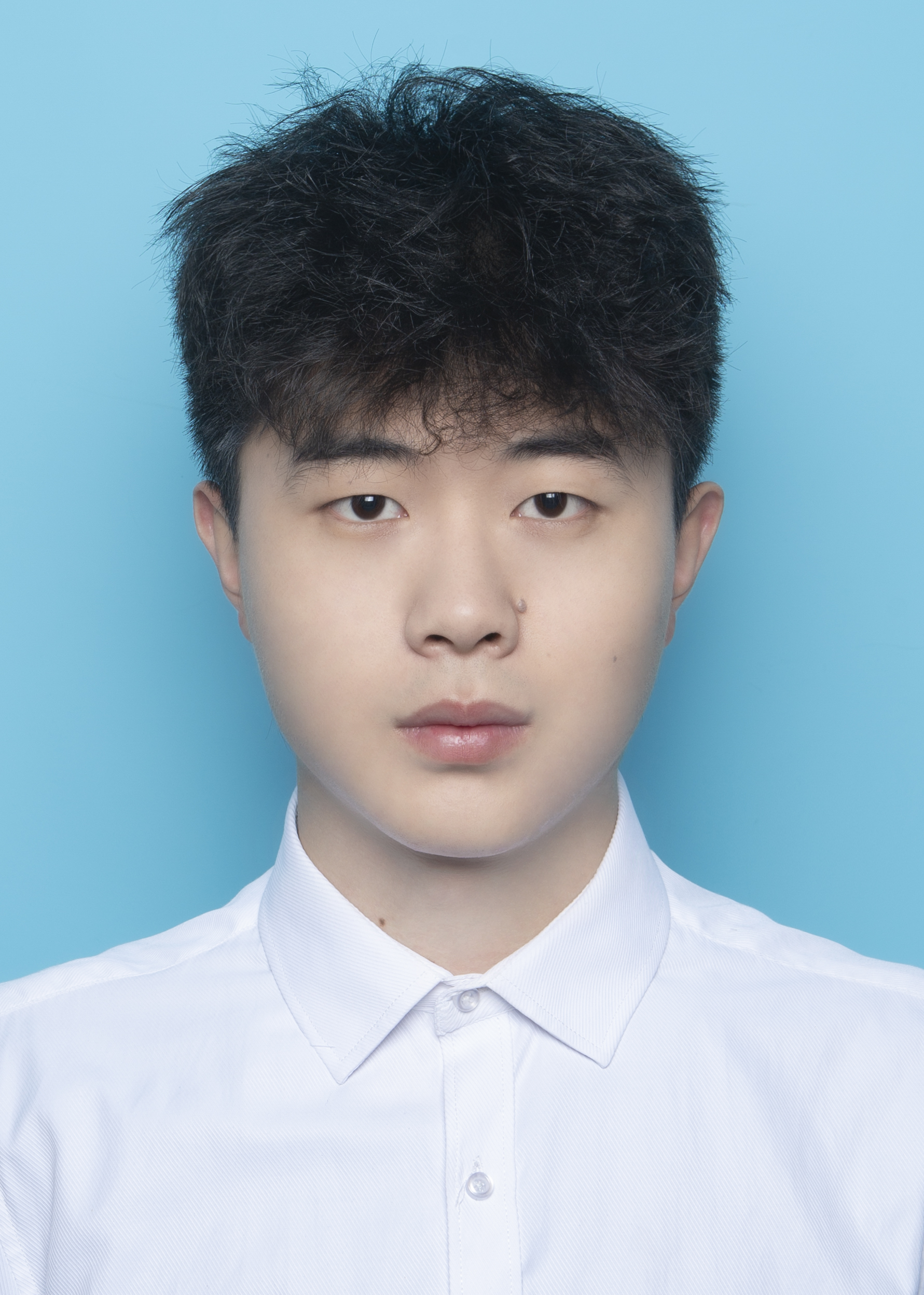}}]{Shengchao Zhu}
received his B.S. degree in Internet of Things Engineering from Hohai University, Changzhou, China, in 2023. He is currently pursuing the Ph.D. degree with the Department of Computer Science and Technology at Hohai University, Nanjing, China. His current research interests include swarm intelligence, swarm ocean, Multi-Agent Reinforcement Learning. 
\end{IEEEbiography}

\vspace{-8ex}

\begin{IEEEbiography}
[{\includegraphics[width=1in,height=1.25in,clip,keepaspectratio]{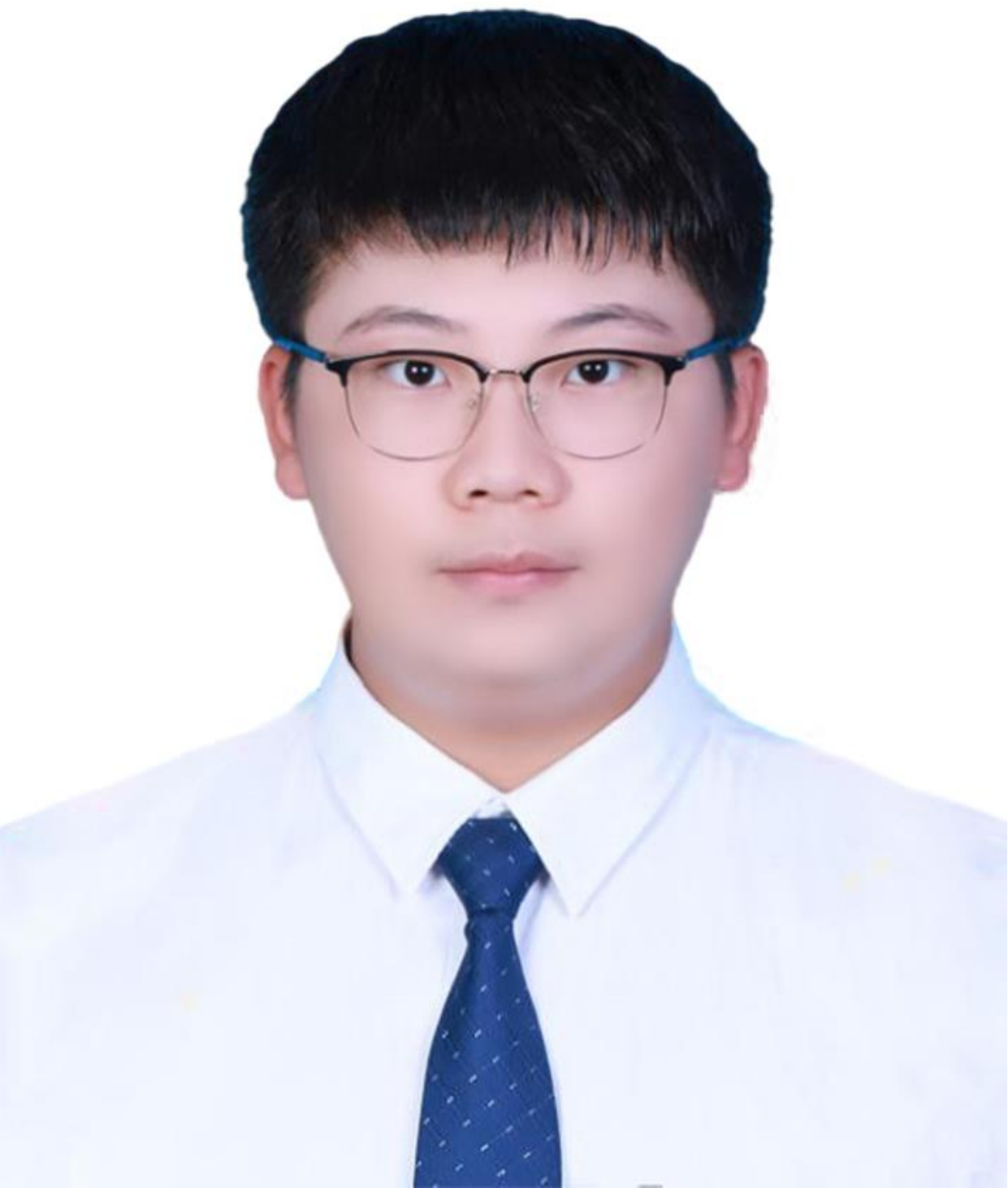}}]{Zhenyu Wang}
is currently pursuing a Bachelor's degree at the Software College, Northeastern University, Shenyang, China. His research interests include reinforcement learning, Internet of things and software-defined networking. 
\end{IEEEbiography}

\end{document}